\documentclass[preprint,review,3p,times]{elsarticle}
\newcommand{\longversion}[1]{#1}
\newcommand{\shortversion}[1]{}

\newcommand{\representchange}[1]{}
\newcommand{\notrepresentchange}[1]{#1}

\newcommand{\amendment}[1]{\notrepresentchange{#1}\representchange{\textcolor{blue}{#1}}}
\newcommand{\clarification}[1]{\notrepresentchange{#1}\representchange{\textcolor{orange}{#1}}}
\newcommand{\reformulation}[1]{\notrepresentchange{#1}\representchange{\textcolor{red}{#1}}}
 \newcommand{\deletion}[1]{\representchange{\textcolor{gray}{\sout{#1}}}}
 
\usepackage[dvipsnames]{xcolor}
\usepackage[normalem]{ulem}
\usepackage{todonotes}
\usepackage{tikz}
\usetikzlibrary{arrows,shapes,snakes,automata,backgrounds,petri}
\usetikzlibrary{shapes.multipart}
\usepackage{comment}
\usepackage[noend]{algpseudocode}
\usepackage{caption}
\usepackage{subcaption}

\usepackage[utf8]{inputenc}          
\usepackage[T1]{fontenc}             
\usepackage[USenglish]{babel}        
\usepackage[intlimits]{amsmath}      
\usepackage{amsfonts}                
\usepackage{amssymb}                 
\usepackage{amsmath}  

\usepackage{hyperref}
\usepackage{cleveref}

\longversion{\usepackage[appendix=inline]{apxproof}}
\shortversion{\usepackage%[appendix=strip]
{apxproof}}

\usepackage{mathtools}
\usepackage{setspace}
\usepackage{graphicx}
\usepackage{hyperref}
\usepackage{dsfont}
\usepackage{xspace}

\usepackage{algorithm}
\usepackage[noend]{algpseudocode}

\usepackage{algorithm}
\usepackage[noend]{algpseudocode}
\usepackage{orcidlink}

\newcommand{\DS}{\textsc{Dominating Set}\xspace}
\newcommand{\EDS}{\longversion{\textsc{Extension Dominating Set}}\shortversion{\textsc{Ext\,DS}}\xspace}
\newcommand{\ERD}{\longversion{\textsc{Extension Roman Domination}}\shortversion{\textsc{Ext\,RD}}\xspace}

\newcommand{\RD}{\longversion{\textsc{Roman Domination}}\shortversion{RD}\xspace}

\newcommand{\HS}{\textsc{Hitting Set}\xspace}
\newcommand{\EHS}{\longversion{\textsc{ Extension Hitting Set}}\shortversion{\textsc{Ext HS}}\xspace}

\newcommand{\RHF}{\shortversion{\textsc{RHF}}\longversion{\textsc{Roman Hitting Function}}\xspace}
\newcommand{\ExtRHF}{\shortversion{\textsc{Ext RHF}}\longversion{\textsc{Extension Roman Hitting Function}}\xspace}
\newcommand{\bERD}{\longversion{Bounded-\textsc{Extension Roman Domination}}\shortversion{Bounded-\textsc{Ext RD}}\xspace}
\newcommand{\mRhfSet}[1]{\mu-\mathcal{RHF} \left(#1\right)}

\newcommand{\NP}{\textsf{NP}\xspace}
\newcommand{\PTIME}{\textsf{P}\xspace}
\newcommand{\FPT}{\textsf{FPT}\xspace}
\newcommand{\XP}{\textsf{XP}\xspace}
\newcommand{\W}[1]{\ensuremath{\textsf{W}[#1]}}
\newcommand{\paraNP}{\textsf{para-NP}\xspace}

\newcommand{\rdf}{\longversion{Roman dominating function}\shortversion{Rdf}\xspace}
\newcommand{\rdfs}{\longversion{Roman dominating functions}\shortversion{Rdfs}\xspace}

\newcommand{\rhf}{\longversion{Roman hitting function}\shortversion{Rhf}\xspace}
\newcommand{\rhfs}{\longversion{Roman hitting functions}\shortversion{Rhfs}\xspace}

\newcommand{\MulIndFam}{\textsc{\longversion{Multicolored Independent Family}\shortversion{MultIndFam}}\xspace}
\newcommand{\hs}{\longversion{hitting set}\shortversion{hs}\xspace}
\newcommand{\ds}{\longversion{dominating set}\shortversion{ds}\xspace}

\newcommand{\no}{\textsf{no}\xspace}
\newcommand{\yes}{\textsf{yes}\xspace}

\newcommand{\Oh}{\mathcal{O}}
\newcommand{\weightfun}{\omega}
\newcommand{\weightfunction}[1]{\weightfun\left(#1\right)}

\newcommand{\iffl}{if\longversion{ and only i}f\xspace}
\newcommand{\swlog}{\longversion{without loss of generality}\shortversion{w.l.o.g}\xspace}

\newtheorem{theorem}{Theorem}
\newtheorem{corollary}{Corollary}
\crefname{corollary}{Corollary}{Corollary}
\newtheorem{lemma}{Lemma}
\crefname{lemma}{Lemma}{Lemma}
\newtheorem*{claim}{Claim}
\crefname{claim}{Claim}{Claim}
\newtheorem{remark}{Remark}
\crefname{remark}{Remark}{Remark}
\newtheorem{proposition}{Proposition}
\crefname{proposition}{Proposition}{Proposition}
\newtheorem{observation}{Observation}
\crefname{observation}{Observation}{Observation}

\newcommand{\reviewer}[1]{\todo[color=red!27]{#1}}\makeatletter
\newenvironment{pf}{\begin{proof}}{\hfill\end{proof}}
\newenvironment{pfclaim}{\begin{proof}}{\renewcommand{\qed}{$\Diamond$}\hfill\end{proof}}
\begin{document}

\begin{frontmatter}
\title{Roman Hitting Functions} %% Author name
\author[1]{Henning Fernau\orcidlink{0000-0002-4444-3220}\fnref{fn1}}
\fntext[fn1]{Corresponding author. Email: fernau@uni-trier.de}
\author[2]{Samuel German\orcidlink{0009-0002-5992-8391}}

\author[1]{Kevin Mann\orcidlink{0000-0002-0880-2513}}
%% Author affiliation
\address[1]{Fachbereich~4 -- Abteilung Informatikwissenschaften, Universit\"at Trier, 54286 Trier, Germany.}
\address[2]{Department of Computer Science and Engineering, University of California San Diego, 9500 Gilman Drive, Mail Code 0404, La Jolla, CA 92093-0404.}

\begin{abstract}
Roman domination formalizes a military strategy dating back to Constantine the Great. Here, armies are placed in different regions. A region is secured if there is at least one army in this region or there are two armies in one neighbored region. This simple strategy can be easily translated into a graph-theoretic question. The placement of armies is described by a function which maps each vertex to~0,~1 or~2. Such a function is called \emph{Roman dominating} if each vertex with value~0 has a neighbor with value~2.

Roman domination is one of few examples where the related (so-called) extension problem is polynomial-time solvable even if the original decision problem is \NP-complete. This is interesting as it allows to establish polynomial-delay enumeration algorithms for listing minimal Roman dominating functions, while it is open for more than four decades if all minimal dominating sets of a graph or (equivalently) if all \amendment{minimal} hitting sets of a hypergraph can be enumerated with polynomial delay, or even in output-polynomial time. To find the reason why this is the case, we combine the idea of hitting set with the idea of Roman domination. We  hence obtain and study a new problem, called \textsc{Roman Hitting Function}, generalizing \textsc{Roman Domination} towards hypergraphs. This allows us to delineate the frontier of polynomial-delay enumerability. 

Our main focus is on the extension version of this problem, as this was the key problem when coping with Roman domination functions. While doing this, we find some conditions under which the \textsc{Extension Roman Hitting Function} problem is \NP-complete. We then use parameterized complexity as a tool to get a better understanding of why \textsc{Extension Roman Hitting Function} behaves in this way. From an alternative perspective, we can say that we use the idea of parameterization to study the question what makes certain enumeration problems that difficult.

Also, we discuss another generalization of \textsc{Extension Roman Domination}, where both a lower and an upper bound on the sought minimal Roman domination function is provided. The additional upper bound makes the problem hard (again), and the applied parameterized complexity analysis (only) provides hardness results. 

Also from the viewpoint of Parameterized Complexity, the studies on extension problems are quite interesting as they provide more and more examples of parameterized problems complete for \W{3}, a complexity class with only very few natural members known some years ago.

We employ our findings to present polynomial-delay and \FPT-delay enumeration algorithms for \rhfs, also drawing the borderline between polynomial delay and transversal hardness.
\end{abstract}

\begin{keyword}enumeration problems, polynomial delay, domination problems, hitting set, Roman domination  
\end{keyword}
\end{frontmatter}

%\newpage

\section{Introduction}
\label{sec:intro}

For more than four decades, the question \amendment{whether} all minimal hitting sets can be enumerated with polynomial delay is an open question.
This \textsc{Transversal Hypergraph Problem} is equivalent to the question if all minimal dominating sets can be enumerated with polynomial delay.\footnote{It is also open if this enumeration is possible in output-polynomial time.}
From the point of view of applications, it is quite important to find an affirmative answer: no user likes to wait ``forever'' to see the next solution, or to get to know that no further solution exist.
In order to explore this question, the problem of enumerating minimal dominating sets has been investigated in many special graph classes.
Several graph classes have been identified where this enumeration problem can be solved with polynomial delay.
In this paper, we approach this problem from a different side.
Namely, in \cite{AbuFerMan2024} it was shown that all minimal Roman dominating functions\footnote{These technical notions will be defined below.} can be enumerated with polynomial delay.
This was a rather surprising finding as in most other complexity aspects, Roman Domination and Dominating Set behave quite the same. More specifically, to mention some of these results:
\begin{itemize}
    \item \textsc{Roman Domination} is \NP-complete, even on special graph classes; see~\cite{Dre2000a,Pagetal2002}. 
    \item \textsc{Minimum Roman Domination} can be approximated up to a logarithmic factor but not any better, unless $\PTIME=\NP$, confer~\cite{AazCheJam2012,Pagetal2002,PadPal2020}.
    \item \textsc{Roman Domination} under standard parameterization (by an upper-bound~$k$ on the number of armies) is complete for \W{2}; see~\cite{Fer08}. However, the dual parameterization puts \textsc{Roman Domination} in \FPT; see \cite{AbuBCF2016,BerFer2015,BerFerSig2014}.\end{itemize}
So, we take this as a basis and try to generalize Roman Domination towards a Hitting Set \amendment{variant} to understand when or where the polynomial delay feature disappears.
In passing, we also generalize the enumeration results from \cite{AbuFerMan2024} considerably. 
As we will explain in the following, we introduce a generalizations of Roman Domination towards hypergraphs; we can describe the tractability frontier (with respect to polynomial delay) quite accurately. 
This also adds to the understanding of the reasons why the Roman variation of domination behaves that differently from the classical setting when it comes to enumeration.

\longversion{Let us first explain why Roman domination variations are interesting by themselves.
Defense strategies of the Roman Empire have been studied throughout the centuries, mostly from a historical-military perspective, but then also from a more graph-theoretic viewpoint, commencing with an article by Arquilla\footnote{Arquilla is quite known for his analysis of cyberwar, see \cite{Arq2011}.} and Fredricksen~\cite{ArqFre95}.
The strategy that we are also discussing in this paper is nowadays known as Roman domination and was introduced in~\cite{Ste99} in line with the discussions of Arquilla and Fredricksen, formalizing a military strategy going back to Constantine the Great.
The concept of Roman domination is a well-studied variation of domination in graphs, 
certified by its own 45-page chapter in the rather recent textbook~\cite{HayHedHen2020}, written by Chellali \emph{et al.}
Here, armies are placed in different regions. A region is secured if there is at least one army in this region or there are two armies in one neighbored region. Mathematically, we model the map as a graph, where regions correspond to vertices and connections between regions correspond to edges; moreover, the placement of armies is described by a function which maps each vertex to~0,~1 or~2. Such a function is called \emph{Roman dominating} if each vertex with value~0 has a neighbor with value~2. \amendment{The pragmatic idea is to pay one to secure a region, but to pay two to also secure all neighboring regions. The goal is to secure the whole empire. }It is clearly desirable to minimize expenses by finding a Roman dominating function that uses the least number of armies; the corresponding decision problem is called \textsc{Roman Domination}.}

Apart from quite a number of combinatorial (graph-theoretic) results that have been obtained for Roman domination, nicely surveyed in\shortversion{ a 45-page chapter of}~\cite{HayHedHen2020}, the decision problem  \textsc{Roman Domination} has been studied from various aspects. 
Even though \textsc{Roman Domination} and \textsc{Dominating Set} behave the same in terms of complexity in a variety of settings this parallelism unexpectedly breaks down for two (mutually related) tasks:
\begin{itemize}
    \item Can we enumerate (i.e., list without repetitions) all minimal solutions of a given instance with polynomial delay?
    \item Can we decide, given a certain part of the solution, if there exists a minimal solution that extends the given pre-solution?
\end{itemize}
The first type of question is also known as an \emph{output-sensitive enumeration} problem. 
Even the less demanding task to enumerate all minimal dominating sets in output-polynomial time is open. \footnote{For a formal definition of some basic notions of output-sensitive enumeration that we only use casually in this discussion, we refer to \cite{GogPapSid98,JohPapYan88a,Str2019}.} 
The corresponding enumeration question for Roman dominating functions \longversion{(for two natural notions of minimality)} can be solved with polynomial delay, as proven in~\cite{AbuFerMan2024} and used in \cite{AbuFerMan2023}.
This result \longversion{on Roman domination }is based on another result giving a polynomial-time algorithm for the \emph{extension problem(s)} as described in the second item. \longversion{Namely, t}\shortversion{T}he idea is to call an extension test before diving further into branching. This strategy is well-established in the area of enumeration algorithms, dating back to Read and Tarjan~\cite{ReaTar75}, but few concrete examples are known; we only refer to the discussion in~\cite{MarStr2019,Str2019}.
This makes \textsc{Extension Roman Domination} one of few examples where the extension problem is polynomial-time solvable, while the original problem is \NP-complete. 
For more details on extension problems, we refer to the survey~\cite{CasFGMS2022}\longversion{ that also suggests a general framework to describe such problems and provides further motivations and applications}. 

The simple scientific question that we want to investigate in this paper is ``why'': \emph{What causes Roman domination to be feasible with respect to enumeration and extension?}
To find out why \textsc{Extension Roman Domination} (\textsc{Ext RD} for short) behaves in this peculiar way and what can be seen as a difference to \textsc{Extension Dominating Set} (\textsc{Ext DS} for short), we are going to generalize the concept of Roman domination and try to find the borderline of tractability. By this, we refer to the question if minimal hitting sets can be enumerated  with polynomial delay. This so-called \textsc{Transversal Hypergraph Problem}~\cite{EitGot95} is open for four decades. It is quite important, as it appears in \amendment{many} application areas, and in particular in databases \cite{CreKPSV2019,EitGot95,GaiVer2017,KanLMN2014}, there are quite a number of interesting equivalent problems, or problems that are shown to be \emph{transversal-hard}, as called in \cite{GogPapSid98}; we only mention two recent references and refer to the papers cited therein: \cite{BlaFLMS2022,BlaFriSch2022}.
 %This question is equivalent to several enumeration problems in logic, database theory and also to enumerating minimal dominating sets in graphs, see \cite{CreKPSV2019,EitGot95,GaiVer2017,KanLMN2014}.
 Our paper can be read as trying to understand which kind of problems are transversal-hard, and furthermore, to describe situations when polynomial-delay enumeration algorithms exist. 
Previous research on enumeration algorithms for minimal dominating sets often tried to look into special graph classes where polynomial-delay enumeration could be exhibited, or not\shortversion{; cf. \cite{KanLMN2011,KanLMN2014} as examples}.\longversion{ For instance, consider split graphs: in~\cite{KanLMN2014}, it is shown that minimal dominating sets can be enumerated with polynomial delay,\footnote{Kanté \emph{et al.} use a direct argument for this polynomial delay enumeration result. It can be also shown that the extension problem to minimal dominating sets is solvable in polynomial time in split graphs.} while minimal connected dominating sets cannot, assuming transversal hardness, see \cite{KanLMN2011}.} This approach can be seen as specializing a known (transversal-hard) enumeration problem by studying special graph classes.
Our approach is different as we come from a domination-type problem with a known polynomial-delay enumeration algorithm for general graphs and we try to stretch this result by generalization to understand when this enumeration task becomes transversal-hard.

It is well-known that \textsc{Hitting Set} (\textsc{HS} for short) can be viewed as a generalization of \textsc{Dominating Set} (\textsc{DS} for short) by modelling a graph by the closed-neighborhood hypergraph. One of the main differences between Dominating Set (in graphs) and Hitting Set is that in the second setting, there is a clear distinction between the objects that can dominate (the vertices of the hypergraph) 
and the objects that should be dominated (which are the hyperedges, i.e., sets of vertices). Although \HS and \textsc{Roman Domination} are both established concepts that generalize \DS, it seems that there is no combination of both concepts published in the literature. Actually, trying to define such a combination comes with some problems.
 If we want a \HS instance to represent a \DS instance by the closed-neighborhood hypergraph, the vertex set of the given graph is the vertex set of the hypergraph, and the set of all closed neighborhoods is the (hyper)edge set. Ignoring twins, this implies a bijection between the universe and the (hyper)edge set. But in general hypergraphs, the number of hyperedges and the number of elements in the universe are independent.  Therefore, we have to think about how to interpret the ``value one setting'' such that it is related to the definition of \textsc{Roman Domination} where exactly one army is put on a certain vertex. We suggest  modelling this effect in hypergraphs based on the following idea: If a vertex has the value~$1$ under a Roman dominating function, then it hits only its ``own'' closed-neighborhood hyperedge. 
In general hypergraphs, we have to explicitly express how a vertex ``owns'' a hyperedge. 
Hence, we need a function, called \emph{correspondence}, which maps a vertex  to an incident hyperedge, such that this hyperedge is dominated if the vertex has the value~1. 
This hypergraph problem is defined more formally in the next section. 

We conclude this section by mentioning that there exist some versions of this paper. First, this paper was a result of splitting the technical report \cite{FerMan2023}. Additional, the technical report included some results of \cite{ManFer2025}. Instead, this paper has on more \W{3}-completeness result (\autoref{thm:ExtRHF-W3} which was a \W{1}-hardness result in \cite{FerMan2023}) as well as more output-sensitive enumeration analysis. This includes transversal-completeness and parameterized enumeration algorithms. An extended abstract of this paper has been presented at IPEC 2024, see \cite{FerMan2024}. This version not only contains full proofs but also a specific new section elaborating the enumerability results that have not been in a special focus at IPEC. \amendment{Moreover, in the conference version, we have raised the following question: \emph{We also do not know if the polynomial-delay enumerability questions that we discussed are really equivalent to the polynomial-delay enumerability of minimal hitting sets.} This question has been settled by Samuel German in the meantime and we have included his arguments (and his name) to this journal version.} 

\section{Definitions and Notation}
\label{sec:definitions}

Throughout this paper, we will freely use standard notions from complexity theory without defining them here. This includes notions from parameterized complexity, concerning \FPT and the further lower levels of the \textsf{W}-hierarchy up to \W{3}, as described in textbooks like~\cite{FluGro2006,DowFel2013}.

\subsection{General Notions, Graphs and Hypergraphs}
\paragraph{General Notions}
Let $\mathbb{N}$ denote the set of all nonnegative integers (including 0). For $n\in \mathbb{N}$, we will use the notation $[n]\coloneqq\{1,\ldots,n\}$. For a finite set~$A$ and some $n\in\mathbb{N}$ with $n\leq |A|$, the cardinality of $A$, $\binom{A}{n}$ denotes the set of all subsets of $A$ of cardinality~$n$, while $2^A$ denotes the power set of~$A$.
For two sets $A,B$, $B^A$ denotes the set of all mappings $f:A\rightarrow B$%\HF{ and $A\times B$ the set $\{(a,b)\mid a\in A,b\in B\}$}\hf{nicht unbedingt nötig}
. If $C\subseteq A$, then $f(C)=\{f(x)\mid x\in C\}\subseteq B$.
%Let $A,B$ be two sets with $A\subseteq B$. 
We denote by $\chi_{C}\in \{0,1\}^A$ the characteristic function, where $\chi_{C}(x)=1$ holds \iffl $x\in C$. For two functions $f,g\in \mathbb{N}^A$, we write $f\leq g$ \iffl $f(a)\leq g(a)$ holds for all $a\in A$. Further, we define the \emph{weight} of~$f$ by $\weightfunction{f}=\sum_{a\in A}f(a)$. 

\paragraph{Hypergraphs}
We focus on not necessarily simple hypergraphs $H=\left(X,\hat{S}=\left(s_i\right)_{i\in I}\right)$ with a finite universe~$X$, also called \emph{vertex} set, and a finite index set $I$, where for each $i\in I$, $s_i\subseteq X$ is a \emph{hyperedge}\footnote{For our proofs, we found this index notation more convenient than a possibly more standard multiset notation.}. With the set $S$ we denote the set  which includes all hyperedges of the family~$\hat{S}$, i.e., $S=\{s_i\mid {i\in I}\}$. Note that the same hyperedge may appear multiple times in the family~$\hat S$. If this is forbidden, we speak of a \emph{simple hypergraph}.\longversion{\footnote{Here, we follow the analogy of \emph{simple graphs} that disallow multiple edges. However, the notion of a \emph{simple hypergraph} was introduced with different meanings before. Ouvrard \cite{Ouv2020} requires the stronger condition that no two hyperedges should include each other. He suggests addressing our simple hypergraphs as hypergraphs without repeated hyperedges. Frieze and Mubayi \cite{FriMub2013} allow two edges of a simple hypergraph to share at most one vertex.}}
 For all $x\in X$, define $\mathbf{S}(x)=\{s_i\in S\mid  x\in s_i\}$ as the set of hyperedges that is \emph{hit} by the vertex~$x$, hence defining a function $\mathbf{S}:X\to 2^S$. A set $D\subseteq X$ is a \emph{hitting set} (\emph{hs} for short) \iffl $\mathbf{S}(D)=S$, where $\mathbf{S}(D)=\bigcup_{x\in D}\mathbf{S}(x)$. Similarly, define $\mathbf{I}:X\to 2^I$ by $\mathbf{I}(x)=\{i\in I\mid x\in s_i\}$, extending to $A\subseteq X$ by $\mathbf{I}(A)=\bigcup_{x\in A}\mathbf{I}(x)$. 
%\todo{it would be nice to give a bit of intuition behind the role of the correspondence function, with a parallel with armies say; a word on how you will consider injective and surjective such functions, together with some intuition on what it means, may be interesting as well.}
We call \longversion{a function }$\tau: X\rightarrow I$ a \emph{correspondence} if $x\in s_{\tau(x)}$ for all $x\in X$ or if, in other words, 
$\tau(x)\in\mathbf{I}(x)$ for all $x\in X$.

Consider a (simple undirected) graph $G=(V,E)$ as a hypergraph $G=(V,\hat{E})$, where each hyperedge contains exactly two elements. %\HF{and $\hat{E}(2^V)\subseteq\{0,1\}$}.\hf{und das hier ...} 
Now, we call the  hyperedges just edges.
Talking about simple graphs, we can consider $E$ as the index set.
For each vertex $v\in V$, define its neighborhood as $N(v)= \{u\mid \{v,u\}\in E\}$ and the closed neighborhood as $N[v] = \{v\} \cup N(v)$. For vertex sets $U\subseteq V$, we use $N[U]= \bigcup_{v\in U} N[v]$ for the closed neighborhood of $U$.  A \emph{dominating set} (\emph{ds} for short) of a graph $G=(V,E)$ is a set $D\subseteq V$ such that $N[D]=V$.

\longversion{In the following, we will formally describe more specific notions concerning Roman Domination and Roman Hitting Functions.

\subsection{Roman Dominating Functions}
Let $G=(V,{E})$ be a graph. }
A function $f:V\to\{0,1,2\}$ is a \emph{Roman dominating function} (Rdf for short) \iffl, for each vertex $v\in V$ with $f(v)=0$, there exists a $u\in N(v)$ with $f(u)=2$. A \rdf $f$ is \emph{minimal} if for each \rdf $g$ with $g\leq f$, $f=g$ holds, leading to the following problems.

\centerline{\fbox{\begin{minipage}{.96\textwidth}
\textbf{Problem name: }\textsc{Roman Domination} (\textsc{RD})\\
\textbf{Given: } A graph $G=(V,{E})$ and $k\in \mathbb{N}$\\
\textbf{Question: } Is there a \rdf $f$ with $\weightfunction{f}\leq k$?
\end{minipage}
}}
\noindent
\centerline{\fbox{\begin{minipage}{.96\textwidth}
\textbf{Problem name: }\textsc{Extension Roman Domination} (\textsc{Ext RD})\\
\textbf{Given: } A graph $G=(V,{E})$ and a function  $f: V\to \{0,1,2\}$\\
\textbf{Question: } Is there a minimal \amendment{\rdf} $g$ for $G$ with $f\leq g$?
\end{minipage}
}}

\clarification{Notice the analogy to \textsc{Dominating Set} and its extension variant, where the input is some set $X$ of vertices and the question is whether there exists a minimal dominating set~$D$ with $X\subseteq D$. For Roman domination, a pointwise ordering of functions is considered, while for domination, the inclusion ordering is taken. This can be reconciled even more by considering characteristic functions instead of sets.}
 Somewhat surprisingly, \reformulation{\textsc{Ext RD}} was proven to be polynomial-time solvable in \cite{AbuFerMan2024}. This implies that minimal \rdf can be enumerated with polynomial delay.
The basis of this algorithm is a combinatorial characterization of minimal \rdfs.
To be able to formulate the\longversion{ already mentioned} combinatorial characterization of minimal \rdf, we need a further notion.
For $D\subseteq V$ and $v\in D$, define the \emph{private neighborhood} of $v\in V$ with respect to~$D$ as $P_{G,D}\left( v\right)\coloneqq N\left[ v\right] \setminus N\left[D\setminus \lbrace v\rbrace\right] $ whose elements are the private neighbors of~$v$.

\begin{theorem} \cite{AbuFerMan2024} \label{t_property_min_rdf}
Let $G=\left(V,E\right)$ be a graph and $f: \: V \to \lbrace 0,1,2\rbrace$ be a function. 
Abbreviate
$G'\coloneqq G\left[ f^{-1}(0)\cup f^{-1}(2)\right]$. Then, $f$ is a minimal \rdf \iffl \shortversion{we find}\longversion{the following constraints hold}:
\begin{enumerate}
\item$N\left[f^{-1}(2)\right]\cap f^{-1}(1)=\emptyset$ \clarification{\emph{(non-interference of 1- and 2-vertices)}},\label{con_1_2}
\item $\forall v\in f^{-1}(2) :\: P_{G',f^{-1}(2)}\left( v \right) \nsubseteq \lbrace v\rbrace$, also called \emph{privacy condition}, and \label{con_private}
\item $f^{-1}(2)$ is a minimal \ds of $G'$ \clarification{\emph{(minimality of  2-vertices)}}.\label{con_min_dom}
\end{enumerate}
\end{theorem}

In order to explore why these results were possible, we \longversion{are going to }generalize these notions and problems for hypergraphs in two ways, with a clear focus on the second possibility.

The first one is probably the most natural one, formed in analogy to the notion of a \ds in a (simple) hypergraph; see \cite{Ach2007}. Let $H=(V,E)$ be a (simple) hypergraph, i.e., $E\subseteq 2^V$. Then, $f:V\to\{0,1,2\}$ is a \rdf if, for all $v\in V$ with $f(v)=0$, there is a vertex $u\in V$ with $f(u)=2$ that is a neighbor of $v$, i.e., it shares an edge with $v$, which means, more formally, that there exists some $e\in E$ with $\{u,v\}\subseteq e$. \longversion{The problems can then be defined as for graphs. We will call them \textsc{Hypergraph Roman Domination}, or  \textsc{HRD} for short, and 
\textsc{Extension Hypergraph Roman Domination}, or  \textsc{Ext HRD} for short.

However, as we show next, we can transfer all interesting properties of Roman domination from the graph case to the hypergraph setting, basically by using the same reduction. More specifically, we can show:
\begin{proposition}
\textsc{HRD} is \NP-complete.
\end{proposition}
\begin{proposition}
\textsc{Ext HRD} can be solved in polynomial time.
\end{proposition}
\begin{proposition}
One can enumerate all minimal \rdfs of a hypergraph with polynomial delay.
\end{proposition}

Namely,}\shortversion{However, as we show next, we can transfer all interesting properties of Roman domination from the graph case to the hypergraph setting by using the same reduction:} if $G=(V,E)$ is a simple hypergraph, we can construct a graph $G'=(V,E')$ by setting
$\{x,y\}\in E'$ \iffl there is a hyperedge $e\in E$ with $\{x,y\}\subseteq e$. This construction is also known as the \emph{Gaifman graph} of~$G$. Then, $f:V\to\{0,1,2\}$ is a \rdf of the hypergraph~$G$ \iffl $f$ is a \rdf of the graph~$G'$.
Conversely, we just have to interpret a given graph as a simple hypergraph.

Therefore, we propose a further generalization of Roman domination towards hypergraphs that allow us to study why Roman domination shows such a peculiar behavior when it comes to its extension version, as well as concerning enumerating minimal \rdfs\shortversion{:

}\longversion{.

\subsection{Roman Hitting Functions}\label{sec_def_rhf}}

\amendment{We now introduce the most important novel concept of this paper.}
Let $H=(X,\hat{S}=(s_i)_{i\in I})$ be a hypergraph and $\tau: X\rightarrow I$ be a correspondence.
\amendment{Coming back to the original motivation behind the concept of Roman domination, 
regions are now modeled by indices of hyperedges, and each region may endorse at most one army. Now, by paying one, an army can secure the endorsed region, and by paying two, one can secure all incident regions.}
 We call a function $f: X\to \{0,1,2\}$ a \emph{Roman hitting function} (Rhf for short) if, for each $i\in I$, there is an $x\in s_i$ with $f(x)=2$ or if there exists an $x\in X$ with $\tau(x)=i$ and $f(x)=1$. In these scenarios, we say that $x$ hits $i$ or $s_i$. For a function $f:X\to \{0,1,2\}$, we define the partition $P(f)=\lbrace 
 f^{-1}(0),f^{-1}(1),f^{-1}(2)
 \rbrace$.
We now define two decision problems related to \rhf.

\centerline{\fbox{\begin{minipage}{.96\textwidth}
\textbf{Problem name: }\textsc{Roman Hitting Function}, or RHF for short\\
\textbf{Given: } A finite set $X$, a hyperedge family $\hat{S}=\left(s_i \right)_{i\in I}$, forming the hypergraph $(X,\hat S)$,
a correspondence $\tau: X\to I$, and $k\in \mathbb{N}$\\
\textbf{Question: } Is there a \rhf $f$ with $\weightfunction{f}\leq k$?\end{minipage}
}}

\noindent
\centerline{\fbox{\begin{minipage}{.96\textwidth}
\textbf{Problem name: }\textsc{Extension Roman Hitting Function}, or \textsc{Ext RHF} for short\\
\textbf{Given: } A finite set $X$, a hyperedge family $\hat{S}=\left(s_i \right)_{i\in I}$, forming the hypergraph $(X,\hat S)$, a correspondence $\tau: X\to I$, and $f: X\to\{0,1,2\}$\\
\textbf{Question: } Is there a minimal \rhf $g$ with $f\leq g$?\end{minipage}
}}

To understand in which way this setting generalizes \RD, recall that there are alternative ways to specify a graph as a hypergraph; namely, the \emph{closed-neighborhood hypergraph} $G_{nb}$ associated to a graph $G=(V,E)$ can be described as $G_{nb}=(V,(N[v])_{v\in V})$. Clearly, $D\subseteq V$ is a \ds \iffl $D$ is a \hs of $G_{nb}$. As $v\in N[v]$, the identity can be viewed as a correspondence.
In this interpretation, $f:V\to \{0,1,2\}$ is a \rdf of~$G$ \iffl it is a \rhf of~$G_{nb}$.

\smallskip
\noindent
\subsection{Organization of the Paper and Main Results}
In \autoref{sec:optimization problems}, we will prove that our optimization problem is \NP-complete and, more interestingly, the problem $k$-$\RHF$ is \W{2}-complete. Then, we turn our attention to the extension problem. Recall that the algorithmic results in the case of Roman domination were based on some basic combinatorial insights\longversion{ that we actually repeat at the end of this section for the reader's convenience}. Following this logic, first we show in \autoref{sec:rhf} a combinatorial characterization of minimal \rhf that we can make use of in  \autoref{sec:tau_sur} where we prove that  \ExtRHF with surjective correspondences can be solved in polynomial time.
In \autoref{sec:bounded-extension-RD}, we return to Roman domination and consider a variant of the extension problem where we give both lower and upper bound conditions to the minimal \rdf that we are looking for. In contrast to the original problem (that only provides a lower bound), this two-sided extension problem turns out to be \NP-complete as we show.%\todo{Ich finde den restlichen Abschnitt doppelt gemoppelt} 
\shortversion{ Furthermore, we identify two natural  parameters under which \bERD is \W{3}-complete. In \autoref{sec:Ext RHF complexity I}, we further discuss different parameterization of \ExtRHF. Again, we obtain some parameterizations for \ExtRHF where the problem is \W{3}-complete.} \longversion{Also a natural parameterization of this problem by the weight of the lower bound function (motivated by the principle of distance to triviality~\cite{GuoHufNie2004}) does not really help, it rather yields a parameterized problem that is complete in the class \W{3} as we prove.
These results are then employed in \autoref{sec:Ext RHF complexity I} to prove that in general, the extension problem for \rhf is \W{3}-complete when parameterized by the weight of the given function. In this section, we investigate alternative parameterizations for this extension problem, sometimes proving hardness results, but sometimes also arriving at \FPT-results. In \autoref{sec:enumerating_rhf}, we show how our results can be applied to develop enumeration algorithms for \rhfs. We focus on an output-sensitive analysis, delineating the borderline between transversal hardness and polynomial delay. We also show that all our enumeration algorithms, they may have polynomial delay or \FPT-delay, use polynomial space only \amendment{(except in the case of \ref{t_converse_transversal_hardness}, which gives a conditionally polynomial-delay enumeration algorithm which isn't necessarily polynomial space)}. 
}\longversion{

Let us mention \longversion{already }here that with our approach, feasibility in the sense of enumerability of minimal solutions with polynomial delay goes hand-in-hand with feasibility in the sense of polynomial-time solvability of the corresponding extension problem. This need not be the case, even in the area of domination-type problems. As shown in \cite{CasFKMS2019a}, the extension problem for edge dominating sets is \NP-complete, while minimal edge dominating sets can be listed with polynomial delay, as shown in~\cite{KanLMNU2015}. Another example related\longversion{ to Roman domination} is exposed in \cite{ManFer2024a}.}%\todo{HF: I commented a paragraph, as a shorter version is in the conclusions.}

\section{The Optimization Problem \textsc{RHF}}
\label{sec:optimization problems}

In this section, we will discuss the (parameterized) complexity of the optimization problem \textsc{RHF}. The probably most natural parameterization for the problem is by an upper bound~$k$ on the weight of the \rhf. For our results, we will use the $\W{2}$-completeness of \RD (with $k$ as parameter) shown in~\cite{Fer08}.
The hardness follows, as \rdfs can be interpreted as \rhfs. For the membership, we construct a split graph, where the clique represents the elements of the universe and the independent set form the hyperedges. Here, the hyperedges are added twice to the independent set if the inverse image of the hyperedge with respect to~$\tau$ is empty. In this way, it is better to put a neighbor to~$2$ than to put these two vertices to $1$.

\begin{theorem}\label{thm:RHF-hardness}
\textsc{RHF} is \NP-complete. The parameterized problem $k$-$\textsc{RHF}$ is \W{2}-complete.
\end{theorem}
\begin{toappendix}
\begin{pf}
As discussed in \autoref{sec:definitions}, \rdfs can be interpreted as \rhfs  so that all hardness claims translate from \RD. 

\clarification{Let us therefore give the membership proofs now.}
Let $H=(X,\hat{S}=(s_i)_{i\in I})$ be a hypergraph, $\tau:X\rightarrow I$ be a correspondence (hence, \amendment{$\tau(x)\in\mathbf{I}(x)$} for each $x\in X$) and $k\in\mathbb{N}$. Define \clarification{the graph} $G=(V,E)$ with
\begin{eqnarray*}
        V &\coloneqq&\, \{a,b,c\} \cup \overbrace{\{v_x\mid x\in X \}}^{=:\,V(X)} \cup\, W\cup U, \quad\text{where}\\
        W &\coloneqq& \{w_{i}\mid  i\in I\}\quad\text{and} \\
        U&\coloneqq& \{u_{i} \mid  i \in I, \tau^{-1}(i)=\emptyset \}\,,\\
        E &\coloneqq& \{\,\{a,b\},\{a,c\}\,\} \cup \{\,\{a,v_x\}\mid x\in X\}\cup \{\,\{v_x,w_{i}\}\mid i\in I, x \in s_i\}\\
        &\cup&\{\,\{v_x,u_{i}\}\mid i\in I, x \in s_i, u_{i} \in U\} \cup \{\,\{v_x,v_y\}\mid x,y\in X\}\,.
\end{eqnarray*}
\clarification{Observe that $G$ is a split graph, with $V(X)\cup\{a\}$ forming a clique and $V\setminus(V(X)\cup\{a\})$ being independent.}
\begin{claim}
$H=(X,\hat{S})$ with \clarification{correspondence} $\tau$ has a \rhf of weight at most $k$ \iffl $G$ has a \rdf with weight at most $k + 2$.
\end{claim}

\begin{pfclaim}
\clarification{We first show the implication from left to right.}
Let $f:X\to \{0,1,2\}$ be \clarification{a} minimal \rhf on~$H$ with \clarification{correspondence}~$\tau$. Define $g: V\rightarrow \{ 0,1,2\}$ \clarification{by setting}
$$ v\mapsto \begin{cases} 2,& \text{if }v\in \{a\}\cup \{v_x \mid  x\in X:\:f(x)=2\}\,, \\ 1,& \text{if }v\in\{w_{i}\in W \mid  \exists x\in \tau^{-1}(i):\:f(x)=1\}\,, \\ 0,& \text{if }v\in \{b,c\}\cup \{w_{i}\in W \mid  \forall x\in \tau^{-1}(i):\:f(x) \neq 1\} \cup U\,.\end{cases}$$

 As $g(a)=2$, $N[a]=\{a,b,c\}\cup V(X)$ is dominated. \clarification{There are two further types of vertices to which $g$ assigns zero. (a)} Let $u_{i}\in U$ for some $ i \in I$. Since $\tau^{-1}(i)=\emptyset$ and as $f$ is a \rhf, there exists an $x\in X\cap s_i$ with $f(x)=2$. \reformulation{By definition,} $v_x\in V(X) \cap N(u_{i})$ with $g(v_x)=2$. Therefore, each vertex in $U$ is dominated. \clarification{(b)} Let $w_{i} \in W$ for some $ i \in I$ with $g(w_{i})=0$. Thus, $f(x)\neq 1$ for each $x\in \tau^{-1}(i)$. As $f$ is a \rhf, there exists an $x\in X\cap s_i$ with $f(x)=2$. Hence, \reformulation{$w_{i}$} is dominated by some $v_x$ and $g$ is a \rdf. \clarification{Finally,}
\begin{equation*}
    \begin{split}
        \sum_{v\in V}g(v)&= 2 + \sum_{ x\in X\text{ with }f(x)=2}g(v_x) + \sum_{i\in I, (\exists x\in \tau^{-1}(i):\:f(x)=1)} g(w_{i})\\
        &\leq 2 + \sum_{x\in f^{-1}(2)}f(x) + \sum_{x\in f^{-1}(1)} f(x) \leq k+2\,.
    \end{split}
\end{equation*}
\clarification{Hence, $g$ is  a \rdf with weight at most $k + 2$.}

For the other direction, assume $f:V\to \{0,1,2\}$ is a \clarification{minimum }\rdf on $G$ \clarification{of weight at most $k+2$}. \clarification{We first derive two combinatorial properties that hold without loss of generality for $f$ as $f$ is minimum.}

 \shortversion{W.l.o.g.}\longversion{\clarification{(A)} Without loss of generality}, if there exist $v,w\in V$ with $N[w]\subseteq N[v]$ and $f(v) + f(w) \geq 2$, then $f(v)=2$ and $f(w)=0$. We can assume this since each other variation \clarification{(i.e., assuming $f(v)=1$ and $f(w)\geq 1$ or $f(v)=0$ and $f(w)=2$)} would dominate at most the same vertices. \clarification{For instance, as $N[w_i]\subseteq N[v_x]$ for $x\in s_i$.} Hence, we \reformulation{can assume that} $\{v\in V\mid f(v)=2\} \subseteq \{a\} \cup \{v_x\mid x\in X \}$ \reformulation{and that $W \cap f^{-1}(2)=\emptyset$}.

\clarification{(B)} \reformulation{Without loss of generality, if there exist vertices $u,v,w\in V$ with $N[u]\cup N[w]\subseteq N[v]$ and $f(u) + f(v) + f(w) \geq 2$, then $f(u)=f(w)=0$ and $f(v)=2$. Namely, let $u,v,w\in V$ with $N[u]\cup N[w]\subseteq N[v]$ and $f(u) + f(v) + f(w) \geq 2$. Assume that $f(v)\neq 2$. If $f(v)=1$, then $f(u) + f(v) + f(w) \geq 2$ brings us back to the proof of Case (A). Hence, we assume  $f(u)=f(w)=1$ and $f(v)=0$.} Define $h:V\to \{ 0,1,2\}$ with $$v\mapsto \begin{cases}f(x),& \text{if }x\notin\{u,v,w\}\,,\\ 0, & \text{if }x\in\{u,w\}\,,\\ 2,& \text{if }x=v\,.\end{cases}$$ $N[u]\cup N[w]$ is dominated (with respect to~$h$) by~$v$. The other vertices are dominated as with \clarification{the \rdf}~$f$\reformulation{. Moreover,}  $\sum_{v\in V}h(v) \leq \sum_{v\in V}f(v)$. Thus, \reformulation{we can assume that} $f(a)=2$ and \reformulation{that} $f(x)=0$ for $x\in \{b,c\} \cup \{w_{i},u_{i}\mid i\in I: \: \tau^{-1}(i) = \emptyset \}$. %Furthermore, $W \cap f^{-1}(2)$ is empty, since we could swap the value of a vertex under $f$ in this set with any of his neighbors. 
For each $w_{i}\in W$ with $f(w_{i})=1$,  \reformulation{$\tau^{-1}(i)\neq \emptyset$}.

\clarification{Now, we define a \rhf on $H$ and $\tau$ based on~$f$.}
Define $g:X\to \{0,1,2\}$ with $g(x)=2$ if $f(v_x)=2$, $g(x)=1$ defined for $x\in \tau^{-1}(i)$  \reformulation{if} there is some $w_{i}\in W$ with $f(w_{i})=1$, and \reformulation{$g(x)=0$}, otherwise. Consider $i\in I$ with no $x\in X$ such that $g(x)=1$ and $\tau(x)=i$ hold. Thus, $f(w_{i})=0$ and there exists a $v_x\in V(X) \cap N(w_{i})$ with $f(v_x)=2$. Hence, there exists an $x\in s_i$ with \reformulation{$g(x)=2$}. Therefore, $g$ is a \rhf. Furthermore, 
$$
        \sum_{x\in X}g(x)= \sum_{x\in X:\: f(v_x)=2} g(x) + \reformulation{\sum_{x\in X:\:\tau(x)=i, f(w_{i})=1} g(x)} =    \sum_{v\in V}f(v)-\underbrace2_{=\,f(a)}\leq k.
$$    
\end{pfclaim}

This concludes the description of the working of our polynomial-time \clarification{many-one} reduction and serves to prove both membership in $\NP$ and membership in \W{2}.
\end{pf}
\end{toappendix}

The fact that \RD is \NP-complete even on split graphs was mentioned repeatedly in the literature, for instance, in \cite{Cocetal2004,Lieetal2008}, but to the best of our knowledge, no proof of this fact has been published. We will provide a strengthened assertion in the following. Recall that two vertices $u,v$ in a graph are called \emph{true twins} if $N[u]=N[v]$.

\begin{lemma}\label{lem:twin-free}
\RD is \NP-complete even on true-twin-free split graphs. Likewise, $k$-\RD is \W{2}-complete on true-twin-free split graphs. 
\end{lemma}
\begin{toappendix}
\begin{pf}
The construction provided in \autoref{thm:RHF-hardness} shows how to construct a graph~$G$ as an instance of \RD, given an instance of \RHF. It is not hard to observe that this graph~$G$ is indeed a split graph as (in the notation of the proof of \autoref{thm:RHF-hardness}) $V(X)\cup\{a\}$ is a clique, while $\{b,c\}\cup W\cup U$ is an independent set.
Hence, any true twins must belong to $V(X)\cup\{a\}$.
As argued in the proof of \autoref{thm:RHF-hardness}, $g(a)=2$ is enforced by $b,c\in N(a)$. \reformulation{This ensures $g(v_x)\neq 1$ as \autoref{t_property_min_rdf} (non-interference of 1- and 2-vertices), would imply the existence of a smaller \rhf.} As $b,c\notin N(V(X))$, there is no $x\in V(X)$ with $N[x]=N[a]$, i.e., true twins must be in $V(X)$. 
\reformulation{Hence, the true-twin relation partitions $V(X)$ into equivalence classes. From each class $C\subseteq V(X)$, choose one representative $v_C$ and let $V'(X)\subseteq V(X)$ be the set of all representatives. Let $G'=G[V'(X)\cup\{a,b,c\}\cup W\cup U\}]$. The graph~$G'$ is  a twin-free split graph that possesses a \rdf of weight at most~$k$ \iffl the graph~$G$ does.}
%now walk through $V(X)$ in an arbitrary but fixed order, possibly deleting  vertices from the graph that are bigger than the currently considered one, and possibly also adding some further vertices to the independent set.
%Let $T(x)=\{y\in V(X)\mid N[y]=N[x]\}$ be the set of twins of~$x$. If $|T(x)|>1$, we delete $T(x)\setminus\{x\}$ from the graph.%, but add vertices $x',x''$ whose only neighbors are~$x$. This enforces setting $g(x)=2$ finally, which is correct, as (in the original graph) setting $g(x)=1$ would mean setting $g(y)=1$ for all twins~$y$ of~$x$.
%The resulting graph gives an instance that is a twin-free split graph that has a \rdf of weight at most~$k$ \iffl the original hypergraph has.
\end{pf}
\end{toappendix}

The hardness part of the proof of \autoref{thm:RHF-hardness} (implicitly) uses the fact that the family of hyperedges could include a hyperedge multiple times as there could be twins (vertices with the same closed neighborhoods) in the original graph. Consider a complete graph $K_n=([n],E_n)$ with $n\geq 2$ vertices. Since the closed neighborhoods are always equal to~$[n]$, if we would use a normal set for the hyperedges instead of a family of hyperedges, then the best solution would be $\chi_{\{v\}}$ for any vertex~$v$. This would even not be a \rdf. Namely, minimal \rdf would be of the form $2\cdot \chi_{\{v\}}$ for any vertex~$v$, or they would be constant~$1$. Nevertheless, the following holds, revisiting \autoref{lem:twin-free}.

\begin{corollary}
\RHF is \NP-complete even on simple hypergraphs. Furthermore, $k$-\RHF is \W{2}-complete.
\end{corollary}

The \NP-completeness of the optimization problem also motivates our analysis of the extension problem; it could help speed up an exact branching algorithm for solving this decision problem.\shortversion{ In the long version, we also consider approximation complexity for \RHF.}\begin{toappendix}
Let us think about \RHF as a minimization problem for some moments.
We mentioned already that \textsc{Minimum RD} can be approximated up to a logarithmic factor. The idea of the proof of \cite{Pagetal2002} can be easily adapted to \RHF and gives us the next result, which we also combine with supplementing hardness results.
\begin{theorem}\label{t_approx_rhf}
There is a polynomial-time  algorithm that, given an instance \reformulation{$(H=(X,(s_i)_{i\in I}),\tau)$} of 
   \textsc{Minimum RHF}, outputs a \rhf $f:X\to \{0,1,2\}$ such that
   $\weightfunction{f}\leq 2\cdot (\ln(|I|)+1)\cdot \text{opt}(H)$.
Moreover, unless $\PTIME=\NP$, there is no polynomial-time algorithm that achieves an approximation ratio better than $\frac{1}{2}(\ln(|I|)-o(1))\cdot \text{opt}(H)$.
\end{theorem}

\begin{pf}
The claimed algorithm would run the well-known greedy algorithm for \HS (or \textsc{Set Cover} in the dual interpretation; see~\cite{Chv79}) on the underlying simple hypergraph and will get a collection $C\subseteq X$ that covers $S=\left\{ s_i\mid i\in I\right\}$. Compared to the optimum cover $C^*$, Chv{\'{a}}tal~\cite{Chv79} proved that $|C|\leq (\ln(|S|)+1)\cdot |C^*|$.
Clearly, $f\coloneqq \chi_C$ is a valid \rhf with $\weightfunction{f}=2\vert C\vert$. Conversely, let $f^*$ be an optimum \rhf solution, i.e., $\weightfunction{f^*}=\text{opt}(H)$. We can obtain a valid solution to the interpretation of $H$ as a \HS instance by defining $C_{f^*}=(f^*)^{-1}(1)\cup (f^*)^{-1}(2)$. Clearly, $|C^*|\leq |C_{f^*}|\leq \weightfunction{f^*}\leq 2\cdot |C_{f^*}|$. Putting the derived inequalities together and finally observing $|S|\leq |I|$ gives the claim on the approximation factor of the described greedy algorithm:
$$\weightfunction{f}=2|C|\leq 2\cdot (\ln(|S|)+1)\cdot |C^*|\leq 2\cdot (\ln(|I|)+1)\cdot \weightfunction{f^*}\,.$$
Now, consider some polynomial-time approximation algorithm~$A$ for \textsc{Minimum RHF} that achieves an approximation ratio of $r(|I|)$, given an instance with index set~$I$. Let $H=(X,S)$ be a simple hypergraph, as an instance of \textsc{Minimum HS} with an optimum \hs $C^*\subseteq X$. We interpret this as an instance of \textsc{Minimum RHF} and run~$A$ on it, getting some solution $f:X\to \{0,1,2\}$. Compared to an optimum solution $f^*$ for this \textsc{Minimum RHF} instance, we know that $\weightfunction{f}\leq r(|S|)\cdot \weightfunction{f^*}$. Let $C_f=f^{-1}(1)\cup f^{-1}(2)$, similar as above. Again, $|C_{f}|\leq \weightfunction{f}\leq 2\cdot |C_{f}|$. If $C_R$ achieves an approximation ratio better than $(\ln(|S|)-o(1))\cdot |C^*|$, then $\PTIME=\NP$ by the famous results of Dinur and Steuer~\cite{DinSte2014}. But this would be the case if $r(|S|)\leq \frac{1}{2}(\ln(|S|)-o(1))$, because then we find:
$$|C_f|\leq r(|S|)\cdot \weightfunction{f^*}\leq r(|S|) \cdot \weightfunction{\chi_{C^*}}\leq r(|S|)\cdot 2|C^*|\leq (\ln(|S|)-o(1))\cdot |C^*|\,,$$
because we could interpret again the \textsc{Minimum HS} solution $C^*$ as a solution $\chi_{C^*}$ of weight $2|C^*|$ that is clearly not better than an optimum \textsc{Minimum RHF} solution~$f^*$.
\end{pf}

There is clearly a small gap between lower and upper bound in the previous theorem, but there is also a gap in the corresponding (in)approximability result concerning \textsc{Minimum Roman Domination} in \cite{Pagetal2002} that is simply not made explicit. However, our reasoning, together with the reductions given at the beginning of this section, prove that also in that case, we still have a multiplicative gap of four. 

Together with the construction given in \autoref{lem:twin-free}, we can deduce the same inapproximability result for \textsc{Minimum RD} on true-twin-free split graphs, hence sharpening previously published results of this type.  

\end{toappendix}

\section{Combinatorial Properties of Minimal Roman Hitting Functions}
\label{sec:rhf}

In this section, we will prove combinatorial properties of minimal \rhfs. This will help us analyze the complexity of \ExtRHF. 

\begin{theorem}\label{t_property_min_rhf}
Let $X$ be a vertex set, $\hat{S}=\left(s_i\right)_{i\in I}$  be a hyperedge family and $\tau:X\rightarrow I$ be a correspondence. Then, a function $f:X\to \{ 0,1,2\}$ is a minimal \rhf \iffl the following constraint items hold:
\begin{enumerate}\setcounter{enumi}{-1}
\item $\forall x,y\in f^{-1}(1) :\: x\neq y \Rightarrow \tau(x) \neq \tau(y)$ \clarification{\emph{($\tau$-injectivity on 1-vertices)}}, \label{con_difvalue}
\item $\forall x\in f^{-1}(1):\: s_{\tau(x)} \cap f^{-1}(2) = \emptyset$ \clarification{\emph{(non-interference of 1- and 2-vertices)}}, \label{con_disjoint}
\item $ \forall x\in f^{-1}(2) \,\exists i\in I\setminus \{\tau(x)\} :\: s_i\cap f^{-1}(2)=\{x\}$ \clarification{\emph{(privacy condition)}}, and \label{con_12switch}
\item $f^{-1}(2)$ is a minimal \hs on $\reformulation{S'\coloneqq{}}\{s_i\in S\mid i\in I, \tau^{-1}(i)\cap f^{-1}(1) = \emptyset\}$ \clarification{\emph{(minimality of 2-vertices)}}. \label{con_minHS}
\end{enumerate}
\end{theorem}

\clarification{The reader is encouraged to compare Constraints~1, 2, 3 with the corresponding items of \autoref{t_property_min_rdf}. This also explains why we start out with item~0 here. These formal similarities are no coincidence; see \autoref{rem:rhf_vs_rdf}.}

%\todo{your characterization gives me the feeling that there is a deep connection between minimal rhf and (not necessarily maximal) irredundant sets in hypergraphs, if not a bijection under some constraints on the correspondence function; if true, this would even more strenghen the interest for rhf and on the other hand explain this difference in complexity between minimal rhd and ds enumerations.}
\begin{pf}
\clarification{Let us first prove the implication from left to right.}
Let $f$ be a minimal \rhf on $X$, $\hat{S}$ and~$\tau$. \clarification{We first show Contraint~\ref{con_difvalue}.} Assume there are $x,y\in f^{-1}(1)$ with $x\neq y$ but $\tau(x)=\tau(y)= i$. Define $\tilde{f}=f-\chi_{\{ y \}}$.  Trivially, $\tilde{f}\leq f$ and $\tilde{f} \neq f$. Since $f^{-1}(2)=\tilde{f}^{-1}(2)$ and $\tau(\tilde{f}^{-1}(1))=\tau(f^{-1}(1)\setminus \lbrace y\rbrace) = \tau(f^{-1}(1))$ hold, $\tilde{f}$ is a \rhf. Thus, $f$ is not minimal, which is a contradiction.

\clarification{Next, we show Constraint~\ref{con_disjoint}.} Assume there \longversion{exists}\shortversion{is} an $x\in f^{-1}(1)$ with $s_{\tau(x)}\cap f^{-1}(2)\neq\emptyset$. Define $\tilde{f}=f - \chi_{\{x\}}$.
Trivially, $ \tilde{f}\leq f$, $f\neq \tilde{f}$ and $f^{-1}(2)=\tilde{f}^{-1}(2)$. Hence, $s_{\tau(x)}\in \mathbf{S}(\tilde{f}^{-1}(2))$.  \reformulation{For $i \in I\setminus \{\tau(x)\}$, $s_i$ is hit by $\tilde{f}$ in the same way as by $f$, so that $\tilde{f}$ is a \rhf.} This contradicts the minimality of~$f$.

\clarification{Finally, we prove Constraint~\ref{con_minHS}.}
Assume $f^{-1}(2)$ is not a minimal hs on $S'$.
If there is an $s\in S'$ that is not hit by $f^{-1}(2)$, then $f$ is no \rhf, contradicting our assumption. 
Hence, we can assume $f^{-1}(2)$ is not minimal. More explicitly,  we assume there is an $x\in f^{-1}(2)$ such that for each $i\in I\setminus \{\tau(x)\}$, $s_i \cap f^{-1}(2)\neq \{x\}$. \clarification{(Notice that this assumption is the negation of Constraint~\ref{con_12switch} that we hence also prove in this argument.)}
Then there is an $x\in f^{-1}(2)$ \reformulation{such that}, for each $s\in \mathbf{S}'(x)\setminus \{\tau(x)\}$, there exists a $y\in \left(f^{-1}(2) \setminus \lbrace x\rbrace\right) \cap s$. Define $\tilde{f}=f-\chi_{\{ x \}}$. Let $i\in I$.
 As $\tau(f^{-1}(1))\cup \{\tau(x)\}= \tau(\tilde{f}^{-1}(1))$ holds by definition, we only need to consider $i\in I\setminus (\tau(f^{-1}(1)) \cup \tau(x))$. For $i$ with $s_i\in \mathbf{S}(f^{-1}(2))\setminus \mathbf{S}(x)$, trivially $s_i\cap \tilde{f}^{-1}(2)\neq\emptyset$. If $s_i\in \mathbf{S}(x)\setminus \{\tau(x)\}$, as we mentioned before, there is a $y\in \left(f^{-1}(2) \setminus \lbrace x\rbrace\right) \cap s_i=\tilde{f}^{-1}(2)\cap s_i$. Thus, $\tilde{f}$ is a \rhf and $f$ is not minimal, contradicting our assumption. Hence, the four conditions hold.

For the if-part, assume $f$ fulfills the constraints. By Constraint~
\ref{con_minHS},\longversion{ we know} for all $i \in I$, either $s_i\cap f^{-1}(2)\neq \emptyset$, or there is an $x\in f^{-1}(1)$ with $\tau(x)=i$. Hence, $f$ is a \rhf. \clarification{We now prove minimality of~$f$.} Let $g:X\to\{0,1,2\}$ be a minimal \rhf with $g\leq f$. Thus, $g^{-1}(2)\subseteq f^{-1}(2)$ and $ \mathbf{S}(g^{-1}(2))\subseteq \mathbf{S}(f^{-1}(2))$ hold. Furthermore, $g^{-1}(1)\subseteq f^{-1}(1)\cup f^{-1}(2)$.  Since for each $x\in X$, $\lbrace s_{\tau(x)}\rbrace \subseteq \mathbf{S}(x)$, for each $i\in \tau(g^{-1}(1))$,  $ i\in \tau(f^{-1}(1))$ or $s_i\in \mathbf{S}(f^{-1}(2))$. \clarification{For the sake of contradiction, let} $x\in X$ be an element with $g(x)< f(x)$. 

%\smallskip
\noindent
\textbf{Case 1:} $g(x)=0<2=f(x)$.
This implies that, for each $i \in \mathbf{I}(x)$, there exists a $y\in s_i$ with $2=g(y)\leq f(y)=2$ or $y\in \tau^{-1}(i)\cap g^{-1}(1) \subset \tau^{-1}(i)\cap (f^{-1}(1)\cup f^{-1}(2))$. This either contradicts Constraint 
\ref{con_disjoint} or Constraint 
\ref{con_12switch}.

%\smallskip
\noindent
\textbf{Case 2:} $g(x)=1<2=f(x)$. This case works analogously, somehow simpler. We only need to exclude $i=\tau(x)$.

\noindent
\textbf{Case 3:} $f(x)=1$.
This implies $g(x)=0$. Since $g$ is a \rhf, either $s_{\tau(x)}\cap g^{-1}(2)$ is not empty or there exists a $y\in g^{-1}(1)$ with $\tau(x)=\tau(y)$.

%\smallskip
\noindent
\textbf{Case 3.1:} $\tau(x)\cap g^{-1}(2)\neq \emptyset$. As $g^{-1}(2)\subseteq f^{-1}(2)$, this contradicts Constraint 
\ref{con_disjoint}.

\noindent
\textbf{Case 3.2:} There is a $y\in g^{-1}(1)$ with $\tau(x)=\tau(y)$. Thus, either there is a $y\in f^{-1}(1)\setminus \lbrace x\rbrace$ \shortversion{with}\longversion{such that} $\tau(x)= \tau(y)$ (this contradicts Constraint 
\ref{con_difvalue}) or $f(y)=2$ (this contradicts Constraint 
\ref{con_disjoint}).

Thus, $g=f$ holds. Therefore, $f$ is minimal.
\end{pf}

\begin{remark}\label{rem:rhf_vs_rdf}
    Now we will compare \autoref{t_property_min_rhf} with \autoref{t_property_min_rdf}. Let $G=(V,E)$ be a graph and let $G_{nb}=(V,(N[v])_{v\in V})$ be the closed-neighborhood hypergraph. The function $\tau: V\to V$ is the identity function. Let $f:V \to \{0,1,2\}$.
    
    As $\tau$ is bijective, Constraint~\ref{con_difvalue} holds for all~$f$. We will show next that $f$ fulfills Constraint~\ref{con_disjoint} of \autoref{t_property_min_rhf} \iffl $f$ fulfills Constraint \ref{con_1_2} of \autoref{t_property_min_rdf}. Assume there is a $v\in f^{-1}(1)$ with $N[v]\cap f^{-1}(2)\neq \emptyset$. Thus $N[f^{-1}(2)]\cap \{v\} \subseteq N[f^{-1}(2)]\cap f^{-1}(1)$ is not empty. This contradicts Constraint~\ref{con_1_2} of \autoref{t_property_min_rdf}. Conversely, the existence of an element $v\in N[f^{-1}(2)]\cap f^{-1}(1)$ would imply that there is a $w\in f^{-1}(2)\cap N[v]$. This contradicts Constraint~\ref{con_disjoint} in \autoref{t_property_min_rhf}.

    For each $v\in f^{-1}(2)$, there is a $u \in V\setminus \{v\}$ with $N[u] \cap f^{-1}(2)=\{v\}$ \iffl each vertex in $f^{-1}(2)$ has a private neighbor that is not itself.  We can reformulate the set occurring in the last Constraint of \autoref{t_property_min_rhf} to $\{N[v]\mid v\in V, \tau^{-1}(v)\cap f^{-1}(1) = \emptyset\} = \{N[v]\mid v\in V,\,  v\notin f^{-1}(1)\}$. This is the set of hyperedges of $(f^{-1}(\{0,2\}), (N[v])_{v\in V\setminus f^{-1}(1) })$. $f^{-1}(2)$ is a minimal hs of this hypergraph \iffl it is a minimal hs of ($(f^{-1}(\{0,2\}), (N[v]\setminus f^{-1}(1))_{v\in V\setminus f^{-1}(1) })$), which is the closed-neighborhood hypergraph of $G'=G[f^{-1}(\{0,2\})]$. Therefore, the last Constraint of \autoref{t_property_min_rdf} is equivalent to the last Constraint of \autoref{t_property_min_rhf} in this case.
\end{remark}

We call a $f:X\to \{0,1,2\}$ \emph{extensible} on the hypergraph $H=(X,\hat{S})$  with correspondence~$\tau$ if there is a minimal \rhf $g$ with $f\leq g$. The following two results are \shortversion{basically }implied by \autoref{t_property_min_rhf}.

\begin{lemma}\label{lem_extensible12}
Let $H=(X,\hat{S}=(s_i)_{i\in I})$ be a hypergraph with correspondence~$\tau$ and $f:X\to \{0,1,2\}$ be a function with $x\in f^{-1}(2)$, $y\in f^{-1}(1)$ and $x\in s_{\tau(y)}$. Then, $f$ is extensible \iffl $f+\chi_{\{y\}}$ is extensible.
\end{lemma}

\begin{toappendix}
\begin{pf}
Since $\leq$ on $\mathbb{N}^X$ is transitive, if $f+\chi_{\{y\}}$ is extensible, then so is~$f$. For the other direction, we assume that there exists a minimal \rhf $g$ with $f\leq g$. Thus, $g(x)=f(x)=2$ and $g(y)\in\{1,2\}$  for $x\in f^{-1}(2)$, $y\in f^{-1}(1)$. By Constraint~\ref{con_disjoint} of \autoref{t_property_min_rhf},  $y\in g^{-1}(2)$, as $x\in s_{\tau(y)}$. For the remaining $z\in X\setminus \{ y \}$, $(f+\chi_{\{y\}})(z)=f(z)\leq g(z)$ holds. This implies $f+\chi_{\{y\}}\leq g$, so that $f+\chi_{\{y\}}$ is extensible. 
\end{pf}
\end{toappendix}

\begin{lemma}\label{lem_extensible11}
Let $H=(X,\hat{S}=(s_i)_{i\in I})$ be a hypergraph with correspondence~$\tau$ and $f:X\to \{0,1,2\}$ be a function with $x, y \in f^{-1}(1)$, $x\neq y$ and $\tau(x) = \tau(y)$. Then, $f$ is extensible \iffl $f+\chi_{\{x,y\}}$ is extensible.
\end{lemma}
\begin{toappendix}
\begin{pf}
Since $\leq$ on $\mathbb{N}^X$ is transitive, if $f+\chi_{\{x,y\}}$ is extensible, then so is~$f$. For the other direction, we assume that there exists a minimal \rhf $g$ with $f\leq g$. Thus, $g(x),g(y)\in\{1,2\}$. By Constraint~\ref{con_difvalue} of \autoref{t_property_min_rhf}, $g(x)=2$ or $g(y)=2$. This implies that if $f$ is extensible, then $f+\chi_{\{x\}}$ or $f+\chi_{\{y\}}$ is extensible. In both cases, we can use \autoref{lem_extensible12} and conclude that $f+\chi_{\{x,y\}}$ is extensible if function~$f$ is extensible.  
\end{pf}
\end{toappendix}

\noindent
\amendment{By use of \autoref{lem_extensible12} and~\autoref{lem_extensible11}, we obtain the following result.   }

\begin{lemma}\label{lem_extensible_equiv}
\amendment{Let $H=(X,\hat{S}=(s_i)_{i\in I})$ be a hypergraph with correspondence~$\tau$ and $f:X\to \{0,1,2\}$ be a function. Then, there exists a function $\widetilde{f}$ for which the following conditions hold:
\begin{enumerate}
    \item There are no two elements $x,y\in X$ with  $f(x)=f(y)=1$ such that $\tau(x)=\tau(y)$.
    \item there are no two elements $x,y\in X$ with $f(x)=2, f(y)=1$ such that $x\in s_{\tau(y)}$.
    \item $f$ is extensible \iffl $\widetilde{f}$ is extensible.
\end{enumerate} }
\end{lemma}
\noindent \clarification{This enables us to formulate a characterization theorem of extensibility that will become quite important later on.}

\begin{theorem}\label{t_extrhf_equiv}
Let $H=(X,\hat{S}=(s_i)_{i\in I})$ be a hypergraph with correspondence~$\tau$, $\tau:X\rightarrow I$. Let $f:X \to \{0,1,2\}$  be a function such that $x\neq y$ implies $\tau(x)\neq \tau(y)$ for each $x,y \in f^{-1}(1)$. Then, $f$ is extensible \iffl there exist both a set $R_2$ with $f^{-1}(2)\subseteq R_2\subseteq f^{-1}(1)\cup f^{-1}(2)$ and a mapping $\rho:R_2\to I$,
%set $\{i_x\}_{x\in R_2}\subseteq I$\hf{meinen wir nicht: $\{i_x\mid x\in R_2\}$; aber die Existenz dieser Menge müssten wir doch gar nicht fordern ...} 
satisfying the following constraints.
\begin{enumerate}
    \item $\forall x\in R_2:\:\rho(x)\neq \tau(x)$.\label{con_tauVSPrivate}
    \item $\forall x\in R_2:\:s_{\rho(x)}\cap R_2= \{x\}$.\label{con_privateEdge}
    \item $\forall x\in f^{-1}(1)\setminus R_2:\: s_{\tau(x)}\cap R_2=\emptyset$.\label{con_1VS2}
    \item $\forall i\in  I \text{ such that } \tau^{-1}(i)=\emptyset: s_i\subseteq\left(\bigcup_{x\in f^{-1}(1)\setminus R_2}s_{\tau(x)}\right) \cup \left( \bigcup_{x\in R_2} s_{\rho(x)}\right)\implies s_i \cap R_2\neq \emptyset$. \label{con_hitting}
\end{enumerate}  
\end{theorem}

\clarification{The idea is that $R_2$ contains those vertices that are ``lifted'' from 1-vertices to 2-vertices as well as the 2-vertices (compared with~$f$) and that $\rho$ then assures privacy for the ``new'' 2-vertices. Constraint \ref{con_hitting} ensures that one can hit all the hyperedges without hitting the private ones.}

\begin{pf}
Define $I'\coloneqq\{i\in I \mid \tau^{-1}(i)=\emptyset\}$. First, we assume that~$f$ is extensible. 
Let $g:X\to \{0,1,2\}$ be a minimal \rhf with $f\leq g$. By Constraint \ref{con_minHS} of \autoref{t_property_min_rhf}, $g^{-1}(2)$ is a minimal \hs on $\{s_i\in S\mid i\in I, \tau^{-1}(i)\cap \clarification{g}^{-1}(1) = \emptyset\}$ $(*)$. With \autoref{t_property_min_rhf}, Constraint \ref{con_12switch}, this implies that, for each $x\in g^{-1}(2)$, there exists an \reformulation{index that we call} $\rho(x) \in I\setminus \{\tau(x)\}$ such that $s_{\rho(x)}\cap g^{-1}(2)=\{x\}$. Define $R_2\coloneqq (f^{-1}(1)\cup f^{-1}(2))\cap g^{-1}(2)$. Clearly, $f^{-1}(2)\subseteq R_2\subseteq f^{-1}(1)\cup f^{-1}(2)$.  \clarification{We can interpret the index function~$\rho$ as a mapping $R_2\to I$.} We have to check the four constraints claimed for~$R_2$ \clarification{and for~$\rho$}. The first \clarification{one is by construction of~$\rho$. The second one is} true in a slightly more general fashion by $(*)$. \clarification{Let us now look at the third constraint.} If there would exist a $y\in f^{-1}(1)\setminus R_2 \subseteq g^{-1}(1)$ with $\emptyset \neq s_{\tau(y)}\cap R_2 \subseteq s_{\tau(y)}\cap g^{-1}(2)$, then this would contradict \autoref{t_property_min_rhf}, Constraint~\ref{con_12switch}. We now turn to the fourth and last constraint. Let $i\in I'$. Since $\tau^{-1}(i)=\emptyset$, there has to be a $y\in g^{-1}(2)\cap s_i$. If $y\in R_2$, then the constraint is satisfied. Hence, we can assume that $y\in g^{-1}(2)\setminus R_2=g^{-1}(2)\cap f^{-1}(0)$.
Consider $x\in f^{-1}(1)\setminus R_2$. As $f\leq g$ and $x\notin R_2$, we have $g(x)=1$. By Constraint~\ref{con_disjoint} of \autoref{t_property_min_rhf}, $y\notin s_{\tau(x)}$.
If $s_i\subseteq\left(\bigcup_{x\in f^{-1}(1)\setminus R_2}s_{\tau(x)}\right) \cup \left( \bigcup_{x\in R_2} s_{\rho(x)}\right)$ and $s_i \cap R_2= \emptyset$ hold, then this would contradict $s_{\rho(x)}\cap R_2 =\{x\}$. Therefore, all the constraints hold.

Let now $f$, $R_2$ and $\rho: R_2\to I$ fulfill the constraints of this theorem. For this part of the proof, we will define a hypergraph $H'$ that includes \reformulation{the edges of $I'$ which are}
 not hit, yet\reformulation{, i.e. there is no $x\in X$ for which $x\in R_2$ or $\tau(X)$ is the edge}. We will show that there is a minimal \hs $D$ on~$H'$ which does not include any vertex of $s_{\rho(x)}$ for $x \in R_2$ or $s_{\tau(x)}$ for $f^{-1}(1)\setminus R_2$. $R_2\cup D$ will describe the set of vertices with value~2. We will hit the remaining vertices by assigning the value 1~to some vertices.
Therefore, we define the hypergraph $H'=(X', (s'_i)_{i\in I''})$ with
\begin{align*}
I''\coloneqq\, &I' \cap \{i\in I \mid s_i\cap R_2 = \emptyset \}\,, \\
X' \coloneqq\, &\left( \bigcup_{i\in I''} s_i \right) \setminus \left( \left(\bigcup_{x\in f^{-1}(1)\setminus R_2}s_{\tau(x)}\right) \cup \left( \bigcup_{x\in R_2} s_{\rho(x)}\right)\right)
\end{align*}
and $s'_i\coloneqq s_i\cap X'$. If $s'_i$ is empty for an $i\in I''$, then there would not exist any \hs on~$H'$. Therefore, we need to ensure that such an index does not exist. Let $i\in I''$. Hence, $\tau^{-1}(i)=\emptyset$ and $s_i\cap R_2 =\emptyset$. The contraposition of the implication of Constraint~\ref{con_hitting} implies
$$s_i\nsubseteq\left(\bigcup_{x\in f^{-1}(1)\setminus R_2}s_{\tau(x)}\right) \cup \left( \bigcup_{x\in R_2} s_{\rho(x)}\right).$$ 
Hence, $s'_i\neq \emptyset$ for each $i\in I''$. Thus, there is a minimal \hs~$D$ on~$H'$. \longversion{Furthermore, t}\shortversion{T}he construction of $H'$ and $D$ implies that $\tau^{-1}(i)\neq \emptyset$ for each $i\in I\setminus \mathbf{I}(R_2 \cup D)$. For each $i \in I\setminus \mathbf{I}(R_2 \cup D)$, $x_i$ will describe an arbitrary vertex in $\tau^{-1}(i)$, unless there exists an $x_i\in \left(f^{-1}(1) \setminus R_2\right) \cap \tau^{-1}(i)$ (by assumption on~$f$, there is at most one such element). In this case, we choose this~$x_i$. 
 
 Define $g:X\to \{0,1,2\}$ with $g^{-1}(1)= \{x_i \mid i \in I\setminus \mathbf{I}(R_2 \cup D)\}$ and $g^{-1}(2)=D\cup R_2$. We will now use \autoref{t_property_min_rhf} to show that $g$ is a minimal \rhf. By the construction of $g^{-1}(1)$, \shortversion{Constraints~\ref{con_difvalue} and~\ref{con_disjoint} of \autoref{t_property_min_rhf}, are fulfilled.}\longversion{the first two constraints of \autoref{t_property_min_rhf}, i.e., Constraints~\ref{con_difvalue} and~\ref{con_disjoint}, are fulfilled.} Since $g^{-1}(1)$ hits each edge in $I\setminus \mathbf{I}(D \cup R_2)$, each hyperedge in $\{s_i \mid i\in I,\tau^{-1}(i)\cap g^{-1}(1)= \emptyset \}$ is hit by $g^{-1}(2)$. As D is minimal and $D\cap \left(\bigcup_{x\in R_2}s_{\rho(x)}\right)=\emptyset$, Constraint \ref{con_privateEdge} implies that is $D\cup R_2$ also a minimal \hs on $\mathbf{I}(D\cup R_2)$. The remaining constraint of \autoref{t_property_min_rhf} follows by the first two constraints of~$f$ together with definition of $H'$ and $D$ as $H'$ only contains hyperedges $s'_i$ where $\tau^{-1}(i)=\emptyset$.\end{pf}
 \shortversion{We will use \autoref{lem_extensible11} and \autoref{t_extrhf_equiv} to show a \W{3}-membership.}
\longversion{
    \begin{remark}
     Let $H=(X, (s_i)_{i\in I})$ be hypergraph with a correspondence $\tau$. \reformulation{By combining} \autoref{lem_extensible11} and \autoref{t_extrhf_equiv}, we can check if a function $f:X\to \{ 0, 1, 2\}$ is extensible (and \longversion{therefore }if $(H,\tau,f)$ is a \reformulation{\yes-}instance of \ExtRHF). First, we use \autoref{lem_extensible11} exhaustively, and then we can use \autoref{t_extrhf_equiv}. \reformulation{This idea} will be used\longversion{ in this paper} to show a \W{3}-membership.
 \end{remark}}
 
\begin{remark}
    \autoref{t_extrhf_equiv} already gives an idea why \ExtRHF with surjective correspondence (and therefore also \ERD) runs in polynomial time. Let $H=(X,(s_i)_{i\in I})$ be a hypergraph and $f:X\to \{0,1,2\}$ be a function with \reformulation{a} surjective correspondence~$\tau$. Hence, we can disregard Constraint \ref{con_hitting}, as $\tau^{-1}(i)\neq \emptyset$ for each $i\in I$. 
    Then, we could use Lemmata~\ref{lem_extensible12} and~\ref{lem_extensible11}. We set $R_2 \coloneqq f^{-1}(2)$.
    By Constraint \ref{con_1VS2}, we have to add each $x\in f^{-1}(1)$ with $s_{\tau(x)} \cap R_2 \neq \emptyset$ to $R_2$. \longversion{This can be done in polynomial time (we will show how in \autoref{alg:tau_sur}). }Now we can check \longversion{in polynomial time }if for each $x \in R_2$ there is a $i_x\in I$ that fulfills Constraints~\ref{con_tauVSPrivate} and~\ref{con_privateEdge}. This will \longversion{also }be our strategy in the next section.
\end{remark}
\section{\ExtRHF with Surjective Correspondence and \ERD}\label{sec:tau_sur}
In this section, we present a polynomial-time algorithm for \ExtRHF instances with a surjective correspondence function~$\tau$. At the end of this section, we explain how this algorithm can be viewed as a natural generalization of \ERD that was studied before in~\cite{AbuFerMan2024}.

\begin{algorithm}[ht]
\caption{ExtRHF Algorithm}\label{alg:tau_sur}
\begin{algorithmic}[1]
\Procedure{ExtRHF Solver}{$X,\hat{S},\tau,f$}\newline
 \textbf{Input:} set $X$, $\hat{S}\coloneqq(s_i)_{i\in I}$,  $\tau$ correspondence function, $f:X\to \{0,1,2\}$ with $\{i\in I \mid \tau^{-1}(i) =  \emptyset \} \subseteq \mathbf{I} \left( f^{-1}(2) \right) $.\newline
 \textbf{Output:} Is there a minimal \rhf $g$ with $f\leq g$?
\State $M_2\coloneqq f^{-1}(2)$\shortversion{,\:}  \label{alg_invariant}
\longversion{\State}$M_1 \coloneqq f^{-1}(1)$
\label{alg_before_while}
\For{$x\in M_1$}\label{alg_tau_not_injective}
\For{$y\in M_1\setminus\lbrace x\rbrace $}
\If{$\tau(x)=\tau(y)$}
\State Add $x,y$ to $M_2$ and delete them from $M_1$.
\State Continue with the next $x$.
\EndIf
\EndFor
\EndFor
\State $ M \coloneqq M_2$ \{ All $x\in g^{-1}(2)$ are considered in the following. \}
\While{$M\neq \emptyset$}\label{alg_whileloop}
\State Choose $x\in M$.
\For {$y\in \tau^{-1}( \mathbf{I}\left( x\right))$}
\If{$y\in M_1$}
\longversion{\State} Add $y$ to $M_2$ and to $M$. Delete $y$ from $M_1$.
\EndIf
\EndFor
\State Delete $x$ from $M$. 
\EndWhile
\For{$x\in M_2$}\label{alg_for_no}
\If{$\mathbf{I}\left( x\right)\setminus \mathbf{I}\left( M_2\setminus \{ x \}\right)\subseteq \{\tau(x)\}$}
\longversion{\State} \textbf{Return \no} \label{alg_no}
\EndIf
\EndFor
\For{$i\in I\setminus \left( \mathbf{I} \left( M_2\right) \cup \tau(M_1)\right)$} \label{alg_fil_for}
\State Add one arbitrary element $x\in \tau^{-1}(i)$ to $M_1 $.
\EndFor 
\State\textbf{Return \yes}  $\{\,g^{-1}(0)=X\setminus \left( M_1\cup M_2\right), g^{-1}(1)=M_1, g^{-1}(2)=M_2\, \}$ 
\EndProcedure
\end{algorithmic}
\end{algorithm}

\begin{theorem}\label{thm:tau_sur}
Algorithm \ref{alg:tau_sur} solves \ExtRHF for instances $(X,\hat{S},\tau,f)$ satisfying $\{i\in I \mid \tau^{-1}(i)= \emptyset \} \subseteq \mathbf{I}\left( f^{-1}(2) \right)$ in polynomial time. 
\end{theorem}
\begin{toappendix}
    \begin{pf}
Assume the algorithm returns \yes. Define $g:X\to\{0,1,2\}$ with $g^{-1}(0)=X\setminus \left( M_1\cup M_2\right)$, $g^{-1}(1)=M_1$ and $ g^{-1}(2)=M_2$. As we  put vertices from~$M_1$ into~$M_2$ and from $X\setminus (M_1\cup M_2)$ into~$M_1$ only, $f\leq g$ holds. Since $\tau^{-1}(i)$ is not empty for each $i\in I\setminus \mathbf{I} \left( M_2\right)\subseteq I\setminus \mathbf{I} \left( f^{-1}(2) \right)$, $g$ is a \rhf. $\tau(M_1)\cap \mathbf{I}(M_2)$ does not contain any elements after the while-loop. As the for-loop starting in line \ref{alg_fil_for} puts elements $x\in X$ in~$M_2$ with $s_{\tau(x)}\cap M_2=\emptyset$ only, $\tau(M_1)\cap \mathbf{I}(M_2)$ is empty at the end of the algorithm. The first for-loop ensures that two different elements in~$M_1$ have different values under~$\tau$. Only in the last for-loop, we put an element~$x$ into~$M_1$, which has a unique $\tau(x)$ and where $\tau(x)$ is not hit so far. If there is a vertex $x\in M_2$ with $\mathbf{I}(x)\setminus \mathbf{I}\left(M_2\setminus \{ x \}\right)\subseteq \{ \tau(x) \}$, then the algorithm would return \no. \autoref{t_property_min_rhf} implies that $g$ is a minimal \rhf.

Assume the algorithm return \no, but there is a minimal \rhf with $f \leq h$. Therefore, the constraints of \autoref{t_property_min_rhf} hold for $h$. This implies that $\mathbf{I}\left(h^{-1}(2)\right) \cap \tau(h^{-1}(1))=\emptyset$ and  $\tau(x) \neq \tau(y)$ hold for each  $x,y\in h^{-1}(1)$ with $x\neq y$. Therefore, $M_1\subseteq h^{-1}(1)$ and $M_2\subseteq h^{-1}(2)$ is fulfilled before line \ref{alg_for_no}. Therefore, $\mathbf{I}(x)\setminus \mathbf{I}\left( h^{-1}(2)\setminus\{x\}\right) \subseteq \mathbf{I}(x)\setminus \mathbf{I}(M_2\setminus\{x\}) =\{\tau(x)\}$ holds for each $x\in M_2$. Thus, $h$ is not a minimal \rdf. Hence, Algorithm~\ref{alg:tau_sur} solves \ExtRHF with $I\setminus \tau(X)\subseteq \mathbf{I}\left(f^{-1}(2)\right)$.  

Trivially, the for-loops run in polynomial time. As in each iteration of the while-loop starting in Line \ref{alg_whileloop} we delete a vertex in $M$ and never put this vertex in to it again, this while-loop needs only polynomial time. Thus, Algorithm~\autoref{alg:tau_sur} runs in polynomial time. 
\end{pf}
\end{toappendix}
\longversion{\noindent
This implies the following corollary. 
\begin{corollary}\label{cor:ExtRHF-surjective}
\ExtRHF with surjective $\tau$ is polynomial-time solvable.
\end{corollary}
}\shortversion{A special case of this theorem entails: \ExtRHF with surjective $\tau$ is polynomial-time solvable.
In the long version, we discuss the connections between \autoref{alg:tau_sur} and Algorithm~1 in~\cite{AbuFerMan2024}.}
\begin{toappendix}
%\begin{remark}
     $G=(V,E)$ be a graph. Recall the closed-neighborhood hypergraph $G_{nb}$  associated to~$G$, i.e., $G_{nb}=(V,(N[v])_{v\in V})$. If we compare \autoref{alg:tau_sur} on $G_{nb}$ with Algorithm~1 in~\cite{AbuFerMan2024} on $G$, then these algorithms would work the same. Since the correspondence~$\tau$ is bijective, $\tau(x)\neq \tau(y)$ holds for each $x,y\in X$ with $x\neq y$. Therefore, the for-loop starting in Line \ref{alg_tau_not_injective} would not change anything. In the while loop we go through all $x\in M_2$ and look for $y\in \tau^{-1}(\mathbf{I}(x))\cap M_1$. The very definition of $\hat{S}$ implies that we update the value of such a $y\in N[x]$ to $f(y)=2$. This also happens in the ExtRDF-algorithm of~\cite{AbuFerMan2024}. In the next step, both algorithms check if each vertex with value 2 has private edge which is not the $\tau$-hyperedge (private neighbor that is not the itself). In the end Algorithm~\ref{alg:tau_sur} takes an index $i$ not hit and sets $f(x)=1$ for one vertex $x\in \tau^{-1}(i)$. As $\tau$ is injective, the vertex~$x$ is unique. Therefore, both algorithm would work the same.   
Our presented algorithm hence generalizes the polynomial-time solvability of \ERD.
%\end{remark}  
\end{toappendix}

\section{Bounded \textsc{Extension Roman Domination}}
\label{sec:bounded-extension-RD}
 
In this section, we will discuss a two-sided bounded version of \textsc{Extension Roman Domination} which was also suggested by a colleague of ours.

\noindent
\centerline{\fbox{\begin{minipage}{.96\textwidth}
\textbf{Problem name: } Bounded \textsc{Extension Roman Domination} (bounded-\textsc{Ext RD})\\
\textbf{Given: } A graph $G=\left( V,E\right)$ and functions $f,h:V\to\lbrace 0,1,2 \rbrace.$\\
\textbf{Question: } Is there a minimal \rdf $g :V\to \lbrace 0,1,2\rbrace$ with $f\leq g \leq h$?
\end{minipage}
}}

\smallskip
We \longversion{will} show in \autoref{cor:bounded-ExtRD-NP} that \bERD is \NP-complete. \longversion{A standard way to act in such circumstances is to}\shortversion{Thus, we} look for \FPT-algorithms.  
One natural parameterization for this problem could be $\weightfunction{2-h}$, because for $2 - h = 0$,\longversion{ i.e., $\weightfunction{2 - h}=0$,} we are back to \ERD as a special case, which is known to be solvable in polynomial time. Hence, this parameterization can be viewed as a `distance-from-triviality' parameter~\cite{GuoHufNie2004}. 
However, as we will prove in \autoref{thm:Bounded ExtRD W3 h0}, this parameterization strategy fails. We employ the well-known \W{3}-completeness of \textsc{Multicolored Independent Family (MultIndFam)}, parameterized by~$k$, in the reduction presented in the proof of \autoref{thm:Bounded ExtRD W3 h0}.

\centerline{\fbox{\begin{minipage}{.96\textwidth}
\textbf{Problem name: }\textsc{Multicolored Independent Family (MultIndFam)}\\
\textbf{Given: } A $(k+1)$-tuple $(S_1,\ldots,S_k,T)$ of subsets of $2^U$
on the common universe $U$.%, i.e., $(U,S_1),\dots,(U,S_k),(U,T)$ are $k+1$ many simple hypergraphs.
\\
\textbf{Question: } Are there hyperedges $s_1\in S_1,\ldots,s_k\in S_k$ such that no $t\in T$ is a subset of $\bigcup_{i=1}^k s_i\subseteq U$?\end{minipage}
}}
In that theorem, we actually discuss a slightly different parameterization, namely $\kappa_{h^{-1}(0)}(G,f,h)\coloneqq\vert h^{-1}(0)\vert$.

\begin{theorem}\label{thm:Bounded ExtRD W3 h0}
$\kappa_{h^{-1}(0)}$-\bERD is \W{3}-hard even on bipartite graphs.
\end{theorem}

\begin{pf}
Let $k\in \mathbb{N}$ and $(S_1,\ldots,S_k,T)$ be a $(k+1)$-tuple of subsets of $2^U$ with a common universe $U$. To simplify the notation, let $T= \{t_1,\ldots,t_{\ell_T}\}$ and $S_i=\{s_{i,1},\ldots s_{i,\ell_i}\}$ for $i\in [k]$. Define $X_j \coloneqq \{x_{j,u}\mid u \in t_j\}$ for $j\in [\ell_T]$, $Y_i \coloneqq \{y_{i,1},\ldots, y_{i,\ell_i}\}$ and $G=(V,E)$ with 
\begin{equation*}
    \begin{split}
        V \coloneqq&\, \left( \bigcup_{i=1}^k \{a_i, b_i, c_i\} \cup Y_i\right) \cup \left( \bigcup_{j=1}^{\ell_T} \{ w_j\} \cup X_j\right)\,,\\
        E \coloneqq&\, \{ \{a_i,y_{i,p}\},\{b_i,y_{i,p}\},\{b_i,c_{i}\} \mid i\in [k],p\in [\ell_i]\} \cup \{ \{w_j,x_{j,u}\} \mid j\in [\ell_T],u\in t_j \} 
        \\  \cup&\, \{ \{ y_{i,p}, x_{j,u} \} \mid i\in [k], p\in [\ell_i], j\in [\ell_T], u\in s_{i,p} \cap t_j \}\,.
    \end{split}
\end{equation*}
\begin{figure}[bt]
    \centering
	\begin{tikzpicture}[transform shape]
		      \tikzset{every node/.style={ fill = white,circle,minimum size=0.3cm}}
            \node  at (-1.3,0)[rectangle]{$\{x_{j,u}\mid u\in s_{i,p}\cap t_j\}$};
            \node [] at (0.6,0){$=$};
			\node[draw,label={right:$w_j$}] (wj) at (3.5,0.75) {};
			\node[draw,label={above:$y_j$}] (yip) at (3.5,-0.75) {};
			\node[draw] (xju1) at (1.5,0.75) {};
			\node[draw] (xju2) at (1.5,-0.75) {};

			\node[draw,label={right:$a_i$}] (ai) at (5.5,0.75) {};
			\node[draw,label={above:$b_i$}] (bi) at (5.5,-0.75) {};
			\node[draw,label={above:$c_i$}] (ci) at (6.5,-0.75) {};
   
            \node at (1.5,0.15) {\vdots};
            \draw (1.5,0) ellipse (15pt and 35pt);	
            \path (yip) edge[-] (xju1);	
            \path (yip) edge[-] (xju2);	
            \path (wj) edge[-] (xju1);	
            \path (wj) edge[-] (xju2);
			\path (yip) edge[-] (ai);
			\path (yip) edge[-] (bi);
			\path (ci) edge[-] (bi);
        \end{tikzpicture}

    \label{fig:g_off_all_TJ_PSpace}
    \caption{Construction for \autoref{thm:Bounded ExtRD W3 h0}, for $i\in [k]$, $p\in [\ell_i]$ and $[j\in \ell_T]$.}
\end{figure}
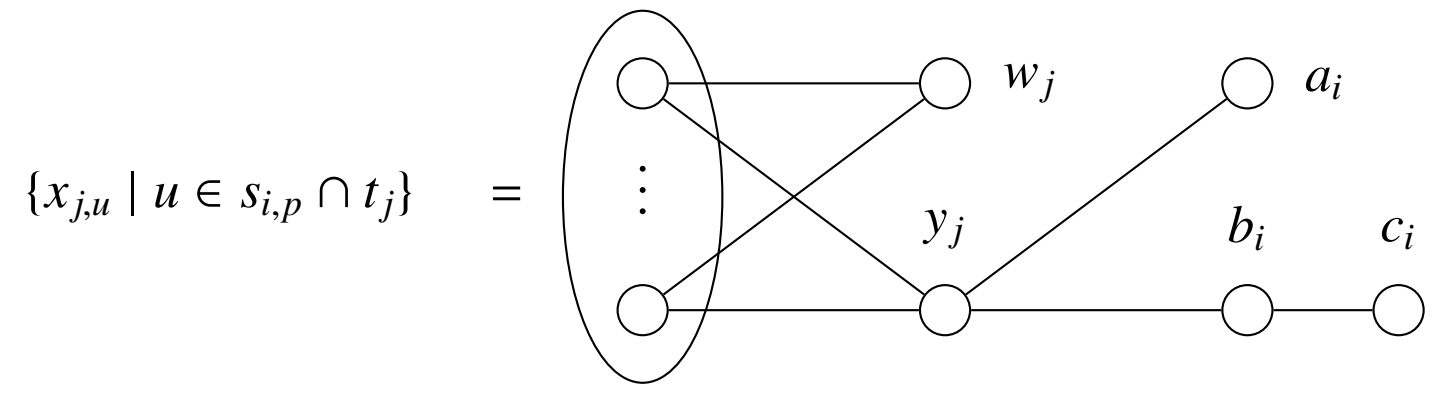
Clearly, $G$ is bipartite, as $V=A\cup B$ decomposes $V$ into two disjoint independent sets, with \begin{itemize}
    \item $A=\left( \bigcup_{i=1}^k \{a_i, b_i\}\right) \cup \left( \bigcup_{j=1}^{\ell_T}  X_j\right)$
and \item $B=\left( \bigcup_{i=1}^k \{c_i\} \cup Y_i\right) \cup \left( \bigcup_{j=1}^{\ell_T} \{ w_j\}\right)$.
\end{itemize}
Furthermore, we need the maps $f,h\in \{0,1,2\}^{V}$ with $f=2\chi_{\{w_1,\ldots,w_{\ell_T}\} \cup \{b_1,\ldots,b_{k}\}}$ and $h=2(1-\chi_{\{a_1,\ldots,a_k\}})$.
\begin{claim}\label{claim:W3equiv hard h0}
    $S_1,\ldots,S_k,T$ is a \yes-instance of the \MulIndFam problem \iffl there exists a minimal \rdf $g$ on $G$ with $f\leq g\leq h$.
\end{claim}
\begin{toappendix}
\begin{pfclaim}\shortversion{[of~\ref{claim:W3equiv hard h0}]}
Let $g$ be a minimal \rdf on $G$ with $f\leq g\leq h$.
Consider $i \in [k]$. As $g(a_i) = h(a_i) = 0$, $Y_i \cap g^{-1}(2)=N(a_i) \cap g^{-1}(2) \neq \emptyset$. Since $N(y_{i,p})\setminus N[g^{-1}(2)] = \{a_i\}$, $\vert Y_i \cap g^{-1}(2)\vert =1$. 
Define for each $i\in [k]$, $p_i\in [\ell_i]$ such that $\{y_{i,p_i}\} =  Y_i \cap g^{-1}(2)$. Assume there is a $j \in [\ell_T]$ such that $t_j\subseteq \bigcup_{i=1}^k s_{i,p_i}$. Then for each $u\in t_j$ there is a $q_u\in [k]$, such that $u\in s_{q_u,p_{q_u}}$ and $x_{j,u}\in N(y_{q_u,p_{q_u}})$. Thus, $w_j$ has no private neighbor but itself. This contradicts to the fact that~$g$ is  minimal. 

For the only-if-part assume $s_{1,d_1}\in S_{1}, \ldots, s_{k,d_k}\in S_{k}$ is a solution to the \MulIndFam\ instance. Define $g \coloneqq f + 2\cdot \chi_{\{y_{1,d_1}, \ldots, y_{k,d_k}\}}$. Clearly, $f\le g\leq h$. As $a_i\in N(y_{i,d_i})$, $Y_i \cup \{ c_i \} = N(b_i)$ for each $i \in [k]$ and $X_j\subseteq w_j$ for $j\in [\ell_T]$, $g^{-1}(0)\subseteq N[f^{-1}(2)]$. Since $g^{-1}(1)=\emptyset$, we only need to show that each vertex in $g^{-1}(2)$ has a private neighbor but itself. 
Clearly, for each $i\in [k]$, $c_i$ is a private neighbor of $b_i$ and $a_i$ of $y_{i,d_i}$. Now assume there is a $j\in [\ell_T]$ such that $w_j$ has no private neighbor but itself. Then for each $x_{j,u}\in N(w_j)$ there is a $q_u\in [k]$ with $x_{j,u} \in N(y_{q_u,d_{q_u}})$ which implies $u\in s_{q_u,d_{q_u}}$. This contradicts $t_j\nsubseteq \bigcup_{i=1}^k s_{i,d_i}$. Hence, $g$ is a minimal \rdf. 
\end{pfclaim}
\end{toappendix}

\noindent
Since  $k = \vert \{a_1, \ldots, a_k\} \vert = \vert h^{-1}(0)\vert$, this is an \FPT-reduction.
\end{pf} 

Since $h$ maps no vertex to~$1$ in the reduction presented in the proof of \autoref{thm:Bounded ExtRD W3 h0}, \bERD is \W{3}-hard, parameterized by $\kappa_{2-h}(G,f,h)\coloneqq\sum_{v\in V}(2-h(v))$. Namely, in the construction  of \autoref{thm:Bounded ExtRD W3 h0}, $\kappa_{2-h}(G,f,h) = 2 \cdot \kappa_{\vert h^{-1}(0) \vert} (G,f,h)$.  

Another parameterization could be $\weightfunction{f}$: If $\weightfunction{f} = 0$, there is a minimal \rdf $g:V\to \{0,1,2\}$ with $f \leq g \leq h$ \iffl $h$ is a \rdf (this can be checked in polynomial time). If there is such a $g$, then $h$ is also a \rdf. If $h$ is a \rdf, then we can decrease the value of the vertices until we can no longer decrease the value of any vertices without losing the \rdf property.    
To understand the complexity of this parameter, we need the following extension version of \HS.

\noindent
\centerline{\fbox{\begin{minipage}{.96\textwidth}
\textbf{Problem name: }\textsc{Extension Hitting Set} (\textsc{Ext HS})\\
\textbf{Given: } A simple hypergraph $H=\left( X,S=(s_i)_{i\in I}\right)$
%, $s_i\subseteq 2^X$,
and a set $U\subseteq X$.\\
\textbf{Question: } Is there a minimal \hs $Y\subseteq X$ with $U \subseteq Y$?
\end{minipage}
}} 

\smallskip
\noindent
In~\cite{BlaFLMS2019}, it was proven that \EHS is \W{3}-complete when parameterized by $|U|$. \shortversion{We use this result in the proof of the following theorem.}

\begin{theorem}\label{thm:Bounded ExtRD W3}
\shortversion{$\weightfun(f)$-\bERD is \W{3}-hard on split graphs.}\longversion{\bERD is \W{3}-hard on split graphs when parameterized by the weight $\weightfun(f)$ of the lower-bound function~$f$. }
\end{theorem}
\begin{toappendix}
\begin{pf}
Let $((X,(s_i)_{i\in I}),U)$ with $X=\{x_1,\ldots,x_n\}$ and $U\neq \emptyset$ be an instance of \textsc{ExtHS}. Define $G=(V,E)$ with 
\begin{equation*}
    \begin{split}
        V &\coloneqq X\cup I,\quad 
        E \coloneqq \{\{x,j\} \mid x_i\in S_j\} \cup \binom{X}{2}.
    \end{split}
\end{equation*}
Obviously, $G$ is a split graph.% Let \reviewer{In the proof of Theorem 8, T and H are already used in the EXT-HS definition above; rename to avoid confusion.} $T\coloneqq\{t_j\mid j\in I\}$%\reviewer{The mapping phi simply renames elements (xi to ci); directly using X would be lighter.}. Define the functions $f,h\in V^{\{0,1,2\}}$ with \shortversion{$f\coloneqq 2\cdot\chi_{\phi(U)}$ and $h \coloneqq \chi_C$}\longversion{$f(t) = f(c) = h(t) = 0$, $f(u) = h(c) = h(u) = 2$, for $u\in \varphi(U), \, t\in T,\, c\in C\setminus \varphi(U)$}. This implies $\sum_{v\in V} f(v)= 2\cdot \vert U\vert.$ Notice that $f(v)=2$ \iffl $\varphi^{-1}(v)\in U$.

Let $Y$ be a minimal \hs with $U\subseteq Y$. Define
\longversion{$$ g_Y: V\rightarrow \{0,1,2\}, x\mapsto \begin{cases}0,& \text{if }x\in I \cup (X\setminus Y)\,,\\
2, & \text{if }x\in Y\,.
\end{cases}$$}\shortversion{$g_Y \coloneqq 2\chi_{\phi(Y)}$.}
$U\subseteq Y$ and $g_Y(T)=\{0\}$ imply $f(v)\leq g_Y(v) \leq h(v)$ for each $v\in V$.
Since $U$ is not empty, $g_Y$ already dominates~$X$. As $Y$ is a \hs, for each $s_j\in S$, there is an $x_i\in S_j\cap Y$. Thus, for each $j\in I$, there is a $x_i\in N(j)$ with $g_Y(x_i)=2$. Hence, $g_Y$ is a \rdf. Assume there exists a $x_i\in X$ with $g_Y(x_i)=2$ such that $g_Y- \chi_{\{x_i\}}$ is also a \rdf. Thus, for each $j\in N(x_i)$ there exists a $x_q\in N(j)\cap g_Y^{-1}(2)$. Hence, for each $s_j\in S$ with $x_i\in s_j$, there is an $x_q\in (s_j\cap Y)\setminus \{x_i\}$. This contradicts the minimality of $Y$. Therefore, $g_Y$ is a minimal \rdf.  Thus,  $(G,f,h)$ is a \yes-instance of \bERD if $(X,S,U)$ is a \yes-instance of \EHS.

Let $g$ be a minimal \rdf on $G$ with $f\leq g \leq h$. Define $Y_g =g^{-1}(2)$%= \{x_i \mid c_i\in C \cap  g(c_i)=2\}$
. $U$ is a subset of $Y_g$ as $f(v)\leq g(v)$ holds for each $v\in V$. Since $ I\subseteq g^{-1}(0)$, there must be a $x_i\in N(j)$ such that $g(x_i)=2$. Thus, $Y_g$ is a \hs. Assume there is an $x_i \in Y_g$ such that for each $ S_j\in S$ with $x_i\in S_j$ there exists an $x_q\in Y_g\subseteq S_j$. Hence, for each $j\in N(x_i) \cap T$, there is a $x_q\in N(j)\setminus\{ c_i \}$ with $g(x_q)=2$. The existence of such a~$x_q$ implies that the private neighbor of $x_i$ cannot be in~$C$. This is a contradiction to the minimality of~$g$.  Thus,  $(X,S,U)$ is a \yes-instance of \EHS  if $(G,f,h)$ is a \yes-instance of \bERD.
\end{pf}
\end{toappendix}

Since the reductions in this section are polynomial-time reductions and since membership in~$\NP$ is easily seen using guess-and-check, we can conclude:

\begin{corollary}\label{cor:bounded-ExtRD-NP}
\bERD is \NP-complete even on split graphs or bipartite graphs.
\end{corollary}
\longversion{\begin{remark}
This result is interesting as we have shown that \EDS is polynomial-time solvable on split graphs \cite{ManFer2025}. As mentioned above, \ERD can be solved in polynomial time but \EDS is \NP-complete. Now we have a graph class for which  \bERD is \NP-complete but there is a polynomial-time algorithm for \EDS.
\end{remark}}

\longversion{Our last proof would also work if the vertex set $T$ would be a clique. This change does not affect the proof as $h(t)=0$ holds for each $t\in T$. The resulting graph is then co-bipartite. Hence, we can state the following results.

\begin{corollary}
    Bounded-\ERD is \NP-complete on co-bipartite graphs. The parameterized problem $\weightfun(f)$-Bounded-\ERD is \W{3}-hard on co-bipartite graphs. 
\end{corollary}}

We will make use\longversion{ of this parameterized reduction proving}\shortversion{ of} the \W{3}-hardness in the next section, when we turn to discuss the\longversion{ (parameterized)} complexity of \ExtRHF. 
The reductions provided there will also show that \bERD is \W{3}-complete for some parameterizations.  \longversion{Thus, the $\W{3}$-hardness results proved in this section turn into $\W{3}$-completeness results.}
\begin{lemma}\label{lem:ExtRHF_trivial_no}
    
\reformulation{Let $(G=(V,E),f,h)$ be a \bERD instance. If one of the there following conditions hold, it is a \no-instance:
    \begin{enumerate}
        \item $f\not\leq h$.
        \item There is a $v\in V$ with $h(v)=0$ and for all $u\in N(v)$, $f(u)\neq 2$.
    \end{enumerate}}
\end{lemma}

\begin{proof}
    \reformulation{\begin{enumerate}
        \item This is trivial, as $\leq$ is transitive. 
        \item Assume there is a $v\in V$ with $h(v)=0$ and for all $u\in N(v)$, $f(u)\neq 2$, $(G,f,h)$ is a no-instance. Then there cannot exist a \rdf $g$ with $f\leq g\leq h$. In particular, if $h(v)=0$, as $g\leq h$, then also $g(v)=0$, and as $g$ should be a \rdf, $g(u)=2$ for some $u\in N(v)$, so that $h(u)=2$ as $g\leq h$.\qed
    \end{enumerate}}
    \renewcommand{\qed}{}
\end{proof}

\section{Complexity of \ExtRHF}
\label{sec:Ext RHF complexity I}
 
%-----------------------------------------------

In this section, we will show that  there are\longversion{ also} instances of \ExtRHF which are\longversion{ \NP-complete and} \W{3}-complete, considering their different parameterizations. For \W{3}-membership, \shortversion{we make again use of \MulIndFam, as there is no further \W{3}-complete problems that we find suitable for a reduction.}\longversion{we make again use of \MulIndFam.
 There are no further \W{3}-complete \longversion{parameterized }problems that we find suitable for a reduction. Hence we use \MulIndFam to prove membership in~\W{3}.}
Unfortunately, this reduction is quite technical.

\begin{theorem}The parameterized problem $\weightfunction{f}$-\ExtRHF is in \W{3}.
\end{theorem}
\begin{pf}
Let  $H=(X,\hat{S}=(s_i)_{i\in I})$ be a \longversion{(not necessarily simple)} hypergraph with correspondence $\tau:X\to I$ and let $f: X\to \{0,1,2\}$ be some function, comprising an instance of $\weightfunction{f}$-\ExtRHF. \reformulation{By \autoref{lem_extensible_equiv} we can assume that there are no two elements $x,y\in X$ such that $f(x)=f(y)=1$ with $\tau(x)=\tau(y)$ or $f(x)=2, f(y)=1$ with $x\in s_{\tau(y)}$. }%\reviewer{This remark should be a lemma if used later. Same comment for the (*) statement in the proof on page 21, lines 7-8 (Das ist diese Stelle).} 
\deletion{Otherwise, we could use Corollaries~\ref{lem_extensible12} and~\ref{lem_extensible11}.}

We will construct an equivalent \MulIndFam instance next. We define its universe as $U\coloneqq X\cup\clarification{\{x'\mid x\in f^{-1}(\{1,2\})\} \cup\{r_{i,x}\mid x\in f^{-1}(\{1,2\}), i\in I\}}\cup \{\tau_x\mid x\in f^{-1}(1)\}\,.$ For the construction of the hypergraphs, we need to define some additional (auxiliary) sets \clarification{whose significance will be explained in the following paragraph}:
\begin{itemize} \item 
For $x \in f^{-1}(\{1,2\})$, $i\in \mathbf{I}(x)$ abbreviate $\tilde{s}_{x,i}\coloneqq s_i\cup\{r_{i,x},x'\}$. 
 \item 
 Define $t_i\coloneqq s_i\cup \{\tau_x\mid x\in f^{-1}(1)\cap s_i\}$ for $i\in I $ \longversion{such that $\tau^{-1}(i)$ and $s_i\cap f^{-1}(2)$ are empty.}\shortversion{with $\emptyset=\tau^{-1}(i)=s_i\cap f^{-1}(2)$.} 
 \item 
For each $x\in f^{-1}(2)$, let $S_x\coloneqq\{\tilde{s}_{x,i}\mid \tau(x) \neq i \}$ and\\ for each $x\in f^{-1}(1)$, let $S_x\coloneqq\{s_{\tau(x)}\cup \{\tau_x\}\}\cup \{\tilde{s}_{x,i}\mid \tau(x) \neq i\}$. 
 \item Furthermore, we need the target set $T=T'\cup T''$, where
\begin{equation*}
    \begin{split}
        T'\coloneqq\,\,&\{\,t_i\mid i\in I\land \tau^{-1}(i)=\emptyset \land s_i \cap f^{-1}(2)=\emptyset\,\}\quad\text{and}\\
     T''\coloneqq\,\,&\left\{\, \{ r_{i,x},y'\}\mid i\in I\land \{x,y\}\subseteq (s_i \cap f^{-1}(\{1,2\}))\land x\neq y \,\right\}\\ 
        &\cup \left\{\,\{\tau_x,y'\} \mid x\in g^{-1}(1)\land y\in  s_{\tau(x)}\land x\neq y \,\right\}\,.
    \end{split}
\end{equation*}
\end{itemize}

Now we will explain the idea of each element. It is important to keep in mind that we want to use \clarification{the extensibility characterization} \autoref{t_extrhf_equiv}: 
If \clarification{element~}$x'$ is in a chosen hyperedge, then we assign the value~2 to~$x$ in the minimal \rhf. The element $r_{i,x}$ gives us information about the mapping $\rho$: $r_{i,x}$ is in one of the chosen edges \iffl $\rho(x)=i$ holds. Therefore, the sets $\{r_{i,x},y'\}$ verify the Constraint~\ref{con_privateEdge} of \autoref{t_extrhf_equiv}. $\tau_x$ will only be in a set we chose if we assign the value~1 to~$x$. Hence, Constraint~\ref{con_1VS2} will be checked by the sets $\{\tau_x ,y\}$. The sets in~$T'$ correspond to the sets which we consider in Constraint~\ref{con_hitting}. This is also the reason why $\tau_x$ \reformulation{is included} in $t_i$. Since there exists a $\tau_x$, $f(x)=1$. If $x\in R_2$, then $s_i\cap R_2\neq \emptyset$. In the \MulIndFam instance, this corresponds to: $\tau_x$ will not be in our sets, which implies that $t_i$ will not be covered completely.   
 
\begin{claim}\label{claim:w3_weight_equiv} 
$(H,\tau,f)$ is a \yes-instance of  \ExtRHF \iffl $(U,(S_x)_{x\in f^{-1}(\{1,2\})},T)$ is a \yes-instance of \longversion{the \MulIndFam problem}\shortversion{\MulIndFam}.
\end{claim}

\begin{toappendix}
\begin{pfclaim}\shortversion{[of~\ref{claim:w3_weight_equiv}]}
First assume that $(H,\tau,f)$ is a \yes-instance of  \ExtRHF, i.e., we have to show that $(U,(S_x)_{x\in f^{-1}(\{1,2\})},T)$ is a \yes-instance of \MulIndFam. 
To prove this, we will use \autoref{t_extrhf_equiv} \reformulation{(after simplifying the instance by using \autoref{lem_extensible_equiv}, as mentioned above)}. As $(H,\tau,f)$ is a \yes-instance of  \ExtRHF, there exist a set $f^{-1}(2)\subseteq R_2 \subseteq f^{-1}(\{ 1,2 \})$ and a mapping $\rho: R_2\to I$ that fulfill all four constraints of \autoref{t_extrhf_equiv}.  For each $x\in R_2$, choose some $\tilde{s}_{x,\rho(x)}\in S_x$, and for each $x \in f^{-1}(1)\setminus R_2$, choose $s_{\tau(x)}\cup \{\tau_x\}\in S_x$. We will show that this choice of hyperedges from $(S_x)_{x\in f^{-1}(\{1,2\})}$ proves that $(U,(S_x)_{x\in f^{-1}(\{1,2\})},T)$ is a \yes-instance of\shortversion{ the problem} \MulIndFam. 
For the sake of contradiction, assume that there exists a hyperedge in~$T$ which is a subset of 
$$\mathcal{S}\coloneqq\left(\bigcup_{x\in R_2}\Tilde{s}_{x,\rho(x)}\right)\cup\left(\bigcup_{x\in f^{-1}(1)\setminus R_2}s_{\tau(x)}\cup \{\tau_x\}\right)\,.$$ 
Observe that, for each $x\in f^{-1}(\{1,2\})$, $x'\in \mathcal{S}$ \iffl there is an $i\in \mathbf{I}(x)$ such that $\tilde{s}_{x,i}$ was chosen. Therefore, $x$ has to be in $R_2$. 

If there would exist a hyperedge in $T''$ which \reformulation{is included} in $\mathcal{S}$, this would contradict  Constraints~\ref{con_tauVSPrivate} or~\ref{con_privateEdge}. Now, consider a hyperedge $t_i$ in $T'$. Hence, we have some $i\in I$ with $t_{i}\subseteq \mathcal{S}$, $s_i\cap f^{-1}(2) = \emptyset$ and $\tau^{-1}(i)=\emptyset$. This implies that $$s_i\subseteq\left(\bigcup_{x\in f^{-1}(1)\setminus R_2}s_{\tau(x)}\right) \cup \left(\bigcup_{x\in R_2} s_{\rho(x)}\right)\,.$$ Furthermore, $ f^{-1}(1) \cap R_2  \cap s_i = \emptyset$, as otherwise, there exists a $\tau_x\in t_i$, which is not in $\mathcal{S}$. This would contradict the Constraint~\ref{con_hitting}. Hence, $(U,(S_x)_{x\in f^{-1}(\{1,2\})}, T)$ is a \yes-instance.

For the if-part, \reformulation{consider the instance $(U,(S_x)_{x\in f^{-1}(\{1,2\})},T)$ of \textsc{MultIndFam}, together with $q_x\in S_x$ for each $x\in f^{-1}(\{1,2\})$ that form a solution for the   instance,} i.e., their union does not contain any hyperedge from~$T$. Define 
$$R_2  \coloneqq f^{-1}(2)\cup \{ x \in f^{-1}(1) \mid \tau_x\notin q_x\}$$
and, for each $x\in R_2$, determine $\rho(x)$ such that $\tilde{s}_{x,\rho(x)}=q_x$. We again use \autoref{t_extrhf_equiv} and the four constraints mentioned in this characterization theorem. Constraint~\ref{con_tauVSPrivate} is fulfilled by the construction of~$R_2$. The next two constraints are met, as otherwise, a set of $T''$ would be included in $\bigcup_{x\in f^{-1}(\{ 1,2 \})} q_x$. Assume there exists a $j\in I$ with $\tau^{-1}(j)=\emptyset$ and 
$$s_j\subseteq \left(\bigcup_{x\in f^{-1}(1)\setminus R_2}s_{\tau(x)}\right) \cup \left( \bigcup_{x\in R_2} s_{\rho(x)}\right)\,.$$
As for each $i\in I$, $t_i$ is no subset of $\bigcup_{x\in f^{-1}(\{1,2\})} q_x$, there must be some $\tau_x\in t_i$, with $ x\in f^{-1}(1)\cap s_j$, which is not in $\bigcup_{x\in f^{-1}(\{1,2\})} q_x$. Thus, $R_2\cap s_i\neq \emptyset.$ Hence, also Constraint~\ref{con_hitting} is met. Therefore, $(H,f,\tau)$ is a \yes-instance  of  \ExtRHF.%\hfill$\Diamond$
\end{pfclaim}
\end{toappendix}
\noindent
As $\vert U\vert \leq \vert X \vert \cdot(\vert I \vert + 4)$, $\vert T \vert  \leq \vert I\vert + \vert X\vert^2 \cdot (\vert I\vert +1)$ and $\vert S_x\vert \leq \vert I\vert$, the $\MulIndFam$ can be constructed in polynomial time. Furthermore, the parameter of the constructed instance of $\MulIndFam$ is $\vert f^{-1}(1)\cup f^{-1}(2) \vert\leq \weightfunction{f}$. 
Hence,  \ExtRHF belongs to $\W{3}$.
\end{pf}

As mentioned in the previous proof, the described reduction is also a poly\-nomial-time reduction. Hence, \ExtRHF is a member of \NP, but this is also \longversion{(easily) }observed by the guess-and-check characterization of \NP.
For the hardness results, we will use \bERD\longversion{ as described in the previous section}.

\begin{theorem}\label{thm:ExtRHF W3}
The parameterized problem 
$\weightfunction{f}$-\ExtRHF is \W{3}-hard %complete 
even if the correspondence function is injective.
Furthermore, $\weightfunction{f}$-\bERD is \W{3}-complete.
\end{theorem}

\begin{pf} We will make use of \autoref{thm:Bounded ExtRD W3}, reducing from \bERD.
Let $(G,f,h)$ be an instance of the \bERD, with $G=(V,E)$. \amendment{By \autoref{lem:ExtRHF_trivial_no}}, we can assume (1) $f\leq h$ and, moreover, (2) $h(v)=0$ implies $h(u)=2$ for some $u\in N(v)$. We parameterize by $\weightfunction{f}$.
For $v\in X \coloneqq V\setminus h^{-1}(0)$, define $T_v\coloneqq\left(N(v) \setminus h^{-1}(\{0,1\})\right)\cup\{v\}$, and  for $v\in h^{-1}(0)$, define $T_v\coloneqq \left(N(v) \setminus h^{-1}(\{0,1\})\right)$. Further, we set
$\hat{S} \coloneqq (T_v)_{v\in V}$ and we define~$\tau$ as the correspondence satisfying $\tau(v)= v$  and we let $\overline{f} :X \rightarrow \{0,1,2\},\, v \mapsto f(v)$, i.e., $\overline{f}=f|_X$. Let $H=(X,\hat{S})$. Then, $(H,\tau,\overline{f})$ describes an instance of \ExtRHF. The parameter is $\weightfunction{\overline{f}}$ for this instance.  As $f\leq h$, $h(v)=0$ implies $f(v)=0$. Thus, $\weightfunction{f}=\weightfunction{\overline{f}}$, so that the parameter value does not change when moving from the \bERD instance to the $\weightfunction{f}$-\ExtRHF instance. Clearly, the described construction can be carried out in polynomial time. Trivially,
\longversion{the correspondence~}$\tau$ is injective. What remains to be shown is the following claim. %construction enjoys the reduction property, mapping \yes-instances to \yes-instances and \no-instances to \no-instances.
\begin{claim}\label{claim:W3hard_weight_equiv}
    $(G,f,h)$ is a \yes-instance of \bERD \iffl $(H,\tau,\overline{f})$ is a \yes-instance of \ExtRHF.\shortversion{\qed}
\end{claim}
\begin{toappendix}
\begin{pfclaim}\shortversion{[of~\ref{claim:W3hard_weight_equiv}]}
Let $g$ be a minimal \rdf on $G$ with $f \leq g \leq h$. Define $\overline{g} :X \rightarrow \{0,1,2\},\, v \mapsto g(v)$, i.e., $\overline{g}=g|_X$. Thus, $\overline{f}\leq \overline{g}$. Let $v\in V$. For $g(v)\in\{1,2\}$, $T_v$ is hit by~$v$. If $g(v)=0$ holds, there has to exist a $w\in N(v)$ with $g(w)=2$. This $w$ hits~$T_v$. Therefore, $\overline{g}$ is a \rhf. 
Assume $\overline{g}$ is not minimal. Since $\tau$ is injective, Constraint~\ref{con_difvalue} of \autoref{t_property_min_rhf} is satisfied. Hence, either Constraint~\ref{con_disjoint}, Constraint~\ref{con_12switch}, or the minimality of Constraint~\ref{con_minHS} of \autoref{t_property_min_rhf} has to be contradicted. If Constraint \ref{con_disjoint} is contradicted, this implies the existence of $v,w\in V$ with $g(v)=\overline{g}(v)=2$, $g(w)=\overline{g}(w)=1$ and $v\in T_w =\left(N(w) \setminus h^{-1}(\{0,1\})\right)\cup\{w\} \subseteq N[w]$. This contradicts the minimality of~$g$ as we can set $g(w)=0$ and still have a \rdf.

Consider Constraint~\ref{con_12switch}. Assume there exists a $v\in \overline{g}^{-1}(2)$ such that for each $w\in V\setminus \{v\}$, $T_w\cap \overline{g}^{-1}(2)\neq \{v\}$ holds. Thus, for each $w\in N(v)$ there exists a $u\in (T_w\cap N[w] \cap \overline{g}^{-1}(2))\setminus \{ v \} \subseteq (N[w] \cap g^{-1}(2))\setminus \{v\}$. This would imply that $g$ is not minimal. 

Concerning Constraint~\ref{con_minHS}, consider $\mathcal{T}\coloneqq\{T_v\mid v\in V, \tau^{-1}(v)\cap \overline{g}^{-1}(1)=\emptyset\}=\{(N(v)\cap h^{-1}(2))\cup\{v\}\mid v\in V,\overline{g}(v)\neq 1\}$. If $\overline{g}(v)=2$, then~$v$ trivially hits $(N(v)\cap h^{-1}(2))\cup\{v\}$. If $\overline{g}(v)=0$, then for some $w\in N(v)$, $g(w)=2$ as $g$ is a \rdf of~$G$. If $g(w)=2$, then $h(w)=2$, so that then $v$ hits $(N(v)\cap h^{-1}(2))\cup\{v\}$. Hence, $\overline{g}^{-1}(2)$ is a \hs of~$\mathcal{T}$. Assume that there exists some $z\in \overline{g}^{-1}(2)$ such that $\overline{g}^{-1}(2)\setminus\{z\}$ is also a \hs of~$\mathcal{T}$. Hence, for each $T_v\in\cal T$ with $z\in T_v$,
there is some~$y\in \overline{g}^{-1}(2)\setminus\{z\}$ with $y\in T_v$. If $v=z$, then $y\in (N(v)\cap h^{-1}(2))$. Hence, $h(y)=2$ and hence $\overline{g}(y)=2$. As $y\in N(z)$ in this case, we could set $g(z)=0$, possibly leading to a smaller \rdf. Now assume $v\neq z$. Then, $z\in (N(v)\cap h^{-1}(2))$. Consider $y\in T_v$. If $y=v$, we can again argue that we may set $g(z)=0$, possibly leading to a smaller \rdf, because we found some $y\in N(z)$ with $g(y)=2$. If $y\neq v$, then $\{y,z\}\subseteq (N(v)\cap h^{-1}(2))$. As $h(v)=2$ implies $g(v)=2$, also in this third case, we can set $g(z)=0$. In summary, in each case, we can find a \rdf for $G$ that is smaller than~$g$ by setting $g(z)=0$. This contradicts the minimality of~$g$. 
By \autoref{t_property_min_rhf}, $\overline{g}$ is minimal.   

Let $\overline{g}$ be a minimal \rhf, a solution to the instance $(H,\tau,\overline{f})$. Define \shortversion{$g:V\to\{0,1,2\}$ with $g(v)=\overline{g}(v)$ for $v\in X$ and $g(v)=0$, otherwise.}
\longversion{$$ g:\: V\rightarrow \{0,1,2\}, v\mapsto \begin{cases} \overline{g}(v), & \text{if }v\in X\text{, i.e., if }h(v)\neq 0\,, \\ 0,& \text{if }v\notin X\text{, i.e., if }h(v)=0\,.\end{cases}$$} 
As $\overline{g}$ extends $\overline{f}$, we have 
$\overline{f}\leq \overline{g}$. By construction, $f\leq h$, so that $f(v)=0$ for $v\notin X$. Hence, $f\leq g$. Consider some $y\in h^{-1}(1)$.
We show \longversion{that then}, $g(y)\leq 1$, which proves $g\leq h$. Since $y\in h^{-1}(1)$, $y$ is only contained in the set $T_y=T_{\tau(y)}$ among all sets in~$\hat S$. If $g(y)=2$, i.e., $\overline{g}(y)=2$,  this would contradict \autoref{t_property_min_rhf}, Constraint~\ref{con_12switch}.
We now prove that $g$ is a \rdf. Consider some  $v\in g^{-1}(0)=\overline{g}^{-1}(0)\cup h^{-1}(0)$.
If $v\in h^{-1}(0)$, there has to exist a $u\in \overline{g}^{-1}(2) \cap T_v \subseteq g^{-1}(2) \cap N[v]$ as there is no $u\in X$ with $\tau(u)=v$.  If $v\in \overline{g}^{-1}(0)$, there is no vertex in $\tau^{-1}(v) \cap \overline{g}^{-1}(1)$ as $\tau(v)=v$. Therefore, there exists a  $u\in \overline{g}^{-1}(2) \cap T_v \subseteq g^{-1}(2)\cap N[v]$.  

Next, we argue why $g$ is minimal. 
Assume there are $v,w\in V$ with $\overline{g}(v)=g(v)=2$, $\overline{g}(w)=g(w)=1$ and $\{v,w\}\in E$, violating the first property of \autoref{t_property_min_rdf}. This would contradict the minimality of $\overline{g}$ as $v\in \left(N(w) \setminus h^{-1}(\{0,1\})\right)\cup\{w\} =T_w$, because  $g(v)=2$ implies $h(v)=2$ by $g\leq h$. Assume that there is a $v\in g^{-1}(2) = \overline{g}^{-1}(2)$ with $N[v]\setminus N[g^{-1}(2) \setminus \{v\}]\subseteq \{v\}$, violating the privacy property of \autoref{t_property_min_rdf}. Thus, for each $w\in N(v)$, there is a $t\in (N[w]\cap \overline{g}^{-1}(2))\subseteq (N(w)\setminus h^{-1}(\{0,1\}))\cup \{ w \}$ with $t\neq v$. 
This contradicts \autoref{t_property_min_rhf} Constraint~\ref{con_12switch}.  Consider
$G'\coloneqq G\left[ g^{-1}(0)\cup g^{-1}(2)\right]$. As~$g$ is a \rdf,  $D\coloneqq g^{-1}(2)$ is a \ds of $G'$. If $D$ is not a minimal \ds (hence violating the last constraint of \autoref{t_property_min_rdf}), then there must be some $v$ with $g(v)=2$ such that $D\setminus \{v\}$ is also a \ds. Hence, for each $u\in N[v]$, there is some $v_u\in N[u]\cap (D\setminus \{v\})$. By construction, for each such $v_u$, we have $g(v_u)=2$ and hence $\overline{g}(v_u)=h(v_u)=2$ and $T_{v_u}\neq \{v_u\}$ (*). On the other hand, as $\overline{g}$ is a minimal \rhf, by the last constraint of \autoref{t_property_min_rhf}, $D=g^{-1}(2)=\overline{g}^{-1}(2)$ is a minimal \hs of $\mathcal{T}\coloneqq\{T_x\mid x\in V\setminus \overline{g}^{-1}(1)\}$. As $T_x\subseteq N[x]$ for all $x\in V$,  $(D\setminus \{v\})$ is not only a  \ds of $G'$ but also a \hs of  $\mathcal{T}$ because of  (*), contradicting mininality as stated in \autoref{t_property_min_rhf}.
As all possible violations of the characterization for minimal \rdf as formulated in of \autoref{t_property_min_rdf} have been shown to contradict our assumptions, we must conclude that $g$ is indeed a minimal \rdf. 
\end{pfclaim}
\end{toappendix}

\longversion{\noindent This concludes our proof.}\shortversion{\renewcommand{\qed}{}\mbox{ }\\[-9ex]}
\end{pf}

\noindent
As this is also a  polynomial-time reduction, it implies following corollary.

\begin{corollary}
\ExtRHF is \NP-complete.
\end{corollary}

We know \longversion{already }that \ExtRHF is polynomial-time solvable if the correspondence function is surjective. \longversion{Since \ExtRHF with injective correspondence function is as hard as the general \HS problem, t}\shortversion{T}his leads to the question if  $\kappa_1(\mathcal{I})=\vert \{i\in I \vert  \tau^{-1}(i)=\emptyset\}\vert$ \longversion{could be}\shortversion{is} a good parameter for this problem for each instance $\mathcal{I}=(H,\tau,f)$ with $H=(V,(s_i)_{i\in I})$, $\tau:V\to I$, $f:V\to\{0,1,2\}$, somehow measuring the \emph{distance from triviality} again. In other words, we try to use parameterized complexity \longversion{as a way }to study the phenomenon that classical function properties as surjectivity seem to be crucial for finding polynomial-time algorithms for \ExtRHF.

\longversion{\begin{theorem}
$\kappa_1$-\ExtRHF is \W{3}-hard.
\end{theorem}

\begin{pf}Observe that the reduction in the proof of \autoref{thm:ExtRHF W3} is also an \FPT-reduction from the parameterized problem $\kappa_{h^{-1}(0)}$-\bERD to $\kappa_1$-\ExtRHF,  employing the parameter $\kappa_{h^{-1}(0)}(G,f,h)=\vert h^{-1}(0)\vert$, so that we can make use of \autoref{thm:Bounded ExtRD W3 h0} towards our claim. 
\end{pf}}

\begin{theorem}\label{thm:ExtRHF-W3}
    $\kappa_1$-\ExtRHF\reformulation{ and $\kappa_{h^{-1}(0)}$-\bERD are} \W{3}-complete.
\end{theorem}
\begin{toappendix}
\begin{pf}
    \shortversion{Observe that the reduction in the proof of \autoref{thm:ExtRHF W3} is also an \FPT-reduction from $\kappa_{h^{-1}(0)}$-\bERD  to $\kappa_1$-\ExtRHF, with $\kappa_{h^{-1}(0)}(G,f,h)\coloneqq\vert h^{-1}(0)\vert$, so that we can employ \autoref{thm:Bounded ExtRD W3 h0} towards our claim.}

    For the membership, we again use the \MulIndFam problem. Let $H=(X,(s_i)_{i\in I})$ be a hypergraph with the correspondence function $\tau: X\rightarrow I$ and two functions $f:X\to\{0,1,2\}$. We assume that \autoref{lem_extensible12} and \autoref{lem_extensible11} used exhaustively on $f$. Define the relation $R=\{(x,y)\in X^2\mid y\in f^{-1}(1), x\in s_{\tau(y)}\}$. Let $R^*$  be the transitive closure and $R_x \coloneqq \{x\} \cup \{y\in X \mid (x,y)\in R_x \}$. We can show by an inductive argument on \autoref{lem_extensible12}, that for any minimal \rhf $g:X\to \{0,1,2\}$ with $f \leq g$ and $g(x)=2$, also $g(y)=2$ for all $y\in R_x$.

    We define $U \coloneqq \{\eta_{v} \mid v \in f^{-1}(\{0,1\})\} \cup \{u_{x,i}\mid x\in X, i\in \mathbf{I}(x) \}$ and $A\coloneqq \{i\in I \mid \tau^{-1}(i)=\emptyset\}$. For $x\in f^{-1}(\{0,1\})$ define $W_x \coloneqq \{u_{y,i} \mid i\in \mathbf{I}(y), y\in s_i\setminus \{x\}\}$ and $r_{x}\coloneqq \bigcup_{y\in R_x}(\{\eta_y\}\cup W_y)$. Furthermore, we need sets $t_x$ for each $x\in X$. For $x \in f^{-1}(2)$, let $t_x \coloneqq \{u_{x,i} \mid i\in \mathbf{I}(x) \setminus \left(\mathbf{I}(f^{-1}(2) \setminus \{ x \}) \cup \{ \tau (x) \}\right)\}$. If $x \notin f^{-1}(2)$, then $t_x\coloneqq \left\{ \eta_x\} \cup \{ u_{x,i}\mid i\in \mathbf{I}(x) \setminus \left(\mathbf{I}(f^{-1}(2)) \cup \{ \tau (x) \}\right) \right\}$. Now we can define the \MulIndFam instance $(S_i)_{i\in A \setminus \mathbf{I}(f^{-1}(2))} \times T$ with $T \coloneqq \{t_x\mid x\in X\}$ and $S_i \coloneqq \{r_x \mid x \in s_i\}$  for $i\in A \setminus  \mathbf{I}(f^{-1}(2))$.
    
    Now we will explain the idea of each element: 
    The set $R_x$ includes each vertex which has to get value~$2$ (by Constraint~\ref{con_1VS2}), if $x$ gets the value~$2$. A vertex~$x$ gets value~1 \iffl $ \eta_x$ is in a chosen hyperedge. If $u_{x,i}$ not in the chosen hyperedges, then $ s_i$ does not contain any element of value~$2$ (besides $x$ probably). Hence, $t_x$ represents the possible private edges of $x$ (beside $\eta_x$). $r_x$ describes the edges, which will covered if $x$ gets the value~$2$ (besides $\eta_y$ possible). 
    \begin{claim}\label{claim:W3_member_RHF}
        $(S_i)_{i\in A \setminus \mathbf{I}(f^{-1}(2))} \times T$ is a \yes-instance of \MulIndFam \iffl $(H,\tau,f)$ is a \yes-instance of \RHF.
    \end{claim}
    \begin{pfclaim}\shortversion{[of \ref{claim:W3_member_RHF}]}
        Let $g$ be a minimal \rhf on $H$ with respect to $\tau$ with $f\leq g$. Hence, for each $i\in A$, there exists an $x_i \in s_i \cap g^{-1}(2)$. For each $i\in A' \coloneqq  A \setminus \mathbf{I}(f^{-1}(2))$, choose $r_{x_i}$ from $S_i$. Assume there is an $x\in X$ with $t_x \subseteq \bigcup_{i\in A'} r_{x_i}$. If $x\in f^{-1}(\{0,1\})$, then $\eta_x \in \bigcup_{i\in A'} r_{x_i}$. Thus, there is a $i\in A'$ such that $(x_i,x)\in R_{x_i}$. Recall, that  $(x_i,x)\in R_{x_i}$ together with $g(x_i)=2$ implies $g(x)=2$. Furthermore, for each $u_{y,i}$ with $i\in \mathbf{I}(y)\setminus(\mathbf{I}(f^{-1}(2)) \cup \{ \tau(y) \})$ there is a $j \in A'$ and a $z\in R_{x_j}$ with $u_{z,i} \in w_z$. This implies, there is no private hyperedge for $x$ under $g$ and $g$ is not minimal. Thus,  $t_x \nsubseteq \bigcup_{i\in A'} r_{x_i}$. This analogously holds for $x\in f^{-1}(2)$. 

        Now let $(r_{x_i})_{i\in A'}$ be solution for the \MulIndFam instance. Define $$B \coloneqq \{x\in A'\mid \eta_x \in \bigcup_{i\in A'}r_{x_i}\}\qquad \text{and}$$ \shortversion{$g \coloneqq f-(2-f)\cdot \chi_B$} \longversion{$$g\colon X\to \{0,1,2\},\quad x\mapsto \begin{cases}
            2, & \text{if }x \in B \,,\\
            f(x), & \text{otherwise}\,.
        \end{cases}$$ }
        We want to show that $P \coloneqq g^{-1}(2) = B \cup f^{-1}(2)$ fulfills the conditions of \autoref{t_extrhf_equiv}. Clearly, $g^{-1}(2)\subseteq P \subseteq g^{-1}(1)\cup g^{-1}(2)$. By definition of $P$, for all $x\in P$, $\eta_x \notin t_x\setminus \bigcup_{i\in A'}r_{x_i}$ if it even exists. Hence, for each $x\in P$ there exists a $i_x \in \mathbf{I}(x) \setminus (I(f^{-1}(2)\setminus \{x\} ) \cup \{\tau(x)\})$ with $u_{x,i_x} \in t_x \setminus \bigcup_{i\in A'}r_{x_i}$. Define now $\rho:B\to I, x\mapsto i_x$. This already implies $\rho(x) \neq \tau(x)$ and $x \in s_{\rho(x)} \cap P$. Assume there is a $y\in  (s_{\rho(x)} \cap P) \setminus \{x\}$. Thus, either $y \in f^{-1}(2)$ ($i_x\notin \mathbf{I}(x) \setminus (I(f^{-1}(2)\setminus \{x\} ) \cup \{\tau(x)\})$) or $\eta_y \in \bigcup_{i\in A'} r_{x_i}$. In the last case $u_{x,i_x}\in w_y \subseteq \bigcup_{i\in A'} r_{x_i}$. Hence, such a $y$ does not exists and $(s_{\rho(x)} \cap P) = \{x\}$.

        Assume there is an $x\in g^{-1}(1)\setminus P=g^{-1}(1)\subseteq f^{-1}(1)$ with $s_{\tau(x)} \cap R_2 = \emptyset$. Then, $\tau(x) \in \mathbf{I}(f^{-1}(2))$ or there is a $y \in s_{\tau(x)}$ with $\eta_y \in  \bigcup_{i\in A'} r_{x_i}$. Hence, either we could use \autoref{lem_extensible12} or $x \in R_y$. In both cases it is a contradiction and such an $x$ does not exists. 
    By definition of $B$ and $A'$, for all $A\subseteq \mathbf{I}(P)$. Hence Condition \ref{con_hitting} is fulfilled and $g$ is  extensible.
    \end{pfclaim} 

\noindent
    As $\vert A'\vert\leq \vert A\vert$, this is an \FPT-reduction.
\end{pf}   
    \end{toappendix}

Now, we consider an explicit \XP-algorithm for $\kappa_1$-\ExtRHF; membership in \XP already follows from \autoref{thm:ExtRHF-W3}.%, $\kappa_{ h^{-1}(0) }$-\bERD, $\kappa_{2-h}$-\bERD

\begin{lemma}\label{lem:ExtRHF_XP}\shortversion{
$\kappa_1$-\ExtRHF, $\kappa_{ h^{-1}(0) }$-\bERD, $\kappa_{2-h}$-\bERD${}\in\XP$.}
\longversion{The following problems are in \XP:
\begin{itemize}
    \item $\kappa_1$-\ExtRHF, \item  $\kappa_{ h^{-1}(0) }$-\bERD, and \item $\kappa_{2-h}$-\bERD.
\end{itemize}}
\end{lemma}
\begin{pf}
For $\kappa_1$-\ExtRHF and $\kappa_{h^{-1}(0) }$-\bERD, it is enough to show that $\kappa_1$-\ExtRHF$\in\XP$. Let $(H,\tau,f)$ be an instance with $H=(X,\hat{S})$, $\hat{S}=(s_i)_{i\in I}$.  This \XP-algorithm is not difficult as we branch over each index in $\{i\in I \mid \tau^{-1}(i)=\emptyset \:\}$ and try to hit the edge $s_i$ with an element from~$X$. If each edge is hit, we check with Algorithm \ref{alg:tau_sur} if there exists a minimal \rhf which is bigger than this function. If Algorithm \ref{alg:tau_sur} returns \yes, we also can return \yes. Otherwise, we try another way to hit all the edges in $S\setminus\tau(X)$.
This algorithm does run in time $\mathcal{O}(n^ {\kappa_1(H,\tau,f)})$.

Now we consider $\kappa_{2-h}$-\bERD. Let $(G,f,h)$ be an instance with $G=(V,E)$ with $f\leq h$. First we want to set $g(x)\coloneqq2$ for each vertex  $x\in N(f^{-1}(2)) \cap f^{-1}(1)$. However, if the set $N(f^{-1}(2)) \cap f^{-1}(1)\cap h^{-1}(\{0,1\})$ is non-empty, we face a trivial \no-instance, as for the solution $g$ we are looking for, any vertex~$x$ with $f(x)=2$ satisfies $g(x)=2$ and moreover, a vertex $y\in N(x)$ with $f(y)=1$ must be set to~$2$ by~$g$, which is ruled out if $h(y)\in\{0,1\}$. As $f\leq g\leq h$, we can set $g(x)\coloneqq 1$ for $x\in f^{-1}(1)\cap h^{-1}(1)$. After these settings, the \XP-algorithm branches on each vertex in $h^{-1}(0)$ and tries to dominate it by a neighbor in $h^{-1}(2)\setminus N(f^{-1}(1)\cap h^{-1}(1))$. As $2 \cdot \vert h^{-1}(0) \vert \leq \weightfunction{2-h}$, in total we have at most $n^{\frac{\weightfunction{2-h}}{2}}$ possibilities to do this. After each of these possible settings for~$g$, where we define $g(x)\coloneqq f(x)$ for any hitherto undefined vertex~$x$, we run the polynomial-time algorithm for \ERD from \cite{AbuFerMan2024}, with $g$ as the lower-bound function. This algorithm will work, as each vertex in $h^{-1}(0)$ is dominated (the undominated vertices are the ones which are set to~$1$) and there is no vertex in $N(f^{-1}(2)) \cap f^{-1}(1)\cap h^{-1}(1)$.
%(The vertices in $N(f^{-1}(2)) \cap f^{-1}(1)$ are these which are set to~$2$).
\end{pf}

We are discussing \longversion{several }other parameterizations\longversion{ in the following} that either lead to  \paraNP-hardness results (\longversion{in }\autoref{t_paraNP_EXTRHF}) or to \FPT-results (\longversion{in }\autoref{t_FPT_EXTRHF}).
%The parameterization function is denoted by $\kappa_Y$, where $Y$ specifies a subset of the universe~$X$ of the instance $(X,\hat S,f)$, such that $\kappa_Y(X,\hat S,f)=\vert Y\vert$.

\begin{theorem}\label{t_paraNP_EXTRHF}
$\kappa_\zeta$-\ExtRHF is \paraNP-hard for each parameterization described by 
$\zeta\in\{ \vert f^{-1}(0)\vert, \vert f^{-1}(1)\vert, \vert f^{-1}(2)\vert, \vert f^{-1}(\{0,2\})\vert \}$.
\end{theorem}

\begin{toappendix}
\begin{pf}
For all the reductions that we show in this proof, we use the \NP-com\-plete\-ness of \EHS. Let $H=(X,\hat S=(s_i)_{i\in I})$ be a hypergraph with $U\subseteq X$, $I\cap X = \emptyset$ and elements $a,i_b,i_c\notin X\cup I$. We define $X'=X \cup \{a\}$, $I'= I\cup X\cup \{i_b, i_c\}$. For each parameterization, we prove that even if the parameter takes values zero or one and is hence fixed, each of the obtained problems can be used to solve arbitrary instances of \EHS.

First we will consider the parameter $\kappa_\zeta$ with $\zeta\in \{f^{-1}(0),f^{-1}(2), f^{-1}(\{0,2\})\}$. 
Define the hypergraph $H'= (X',(s_i')_{i\in I'})$ and the correspondence $\tau$ with 
$$ \tau: X'\to I', x\mapsto \begin{cases}
    x, & \text{if }x\in X\,.\\
    i_b, & \text{if }x=a\,,
\end{cases} \qquad s_i'\coloneqq\begin{cases}
    s_i, & \text{if }i\in I\,,\\
    \{a,i\}, & \text{if }i\in U\,,\\
    \{ i \}, & \text{if }i\in X\setminus U\,,\\
    X', & \text{if }i=i_b\,,\\
    \{ a \}, & \text{if }i=i_c\,,\\
\end{cases}$$ for $i\in I'$. Furthermore, we need \shortversion{$f=2\cdot\chi_{\{ a \}}+ \chi_X \in\{0,1,2\}^{X'} $}
\longversion{$$ f: X'\to \{0,1,2\}, x \mapsto \begin{cases}
    2, & \text{if }x=a\,,\\
    1, & \text{if }x \in X\,.
\end{cases}$$ }This implies $\vert f^{-1}(\{0,2\})\vert =1 = \vert f^{-1}(2)\vert$ and $\vert f^{-1}(0)\vert =0 $, i.e., the parameter values are constant.
\begin{claim}\label{claim:paraNPf02}
    $(H,U)$ is a \yes-instance of \EHS \iffl $(H',\tau,f)$ is a \yes-instance of \ExtRHF.
\end{claim}
\begin{pfclaim}
First, we use \autoref{lem_extensible12} exhaustively. As $f(a)=2$ and $a\in s_{\tau(x)}$ for $x\in U$, $(H',\tau,f)$ is a \yes-instance of \ExtRHF \iffl  $(H',\tau,\tilde{f})$ with $\tilde{f}(x)=2$ for $x\in U\cup \{a\}$ and $\tilde{f}(x)=1$ for $x\in X \setminus U$ is a \yes-instance of \ExtRHF. 
Assume that there exists a minimal \rhf~$g$ of $(H',\tau)$ %$H'$, $\tau$ 
with $\tilde{f} \leq g$. Define $D= g^{-1}(2)\setminus \{a\}$. As $a\notin s_i'=s_i$ for each $i\in I$ (i.e., $\tau^{-1}(i)=\emptyset$), $s_i\cap D\neq \emptyset$  for such an~$i$. If there existed an $x\in D$  with for each $i\in \mathbf{I}(x)$ there is a $y\in s_i\cap (D\setminus \{x\})$, then this would contradict Constraint \ref{con_12switch} of \autoref{t_property_min_rhf}. Hence, $D$ is a minimal \hs.  

Assume that there is a minimal \hs $D\subseteq X$ on $H$ with $U\subseteq D$. Define $g\in \{0,1,2\}^{X'}$ with $g(x)=2$ for $x\in D\cup \{ a \}$ and $g(x)=1$ for $x\in X\setminus D$. We use \autoref{t_property_min_rhf} to show that $g$ is a minimal \rhf. Since $s_{\tau(x)}=\{ x \}$ for each $x\in g^{-1}(1)= X\setminus D\subseteq X\setminus U$, the two first constraints hold. The fact that $D$ is a minimal \hs on~$H$ and $s_{i_c}\cap g^{-1}(2)= \{a\}$ imply Constraint \ref{con_12switch}. Constraint~\ref{con_minHS} follows, because $D$ is a \hs on $H$ and the hyperedges $(s_i')_{i \in I'\setminus I}$ are hit by $\{a\}\cup (X\setminus D)$. Therefore, $g$ is a minimal \rhf of~$H'$. 
\end{pfclaim}
%----------------------------------------------------------------------------------------
%----------------------------------------------------------------------------------------
    
    \shortversion{W}\longversion{Next, w}e take a look at the parameter described by $\vert f^{-1}(1)\vert$. Define the hypergraph $H''= (X',(s_i'')_{i\in I'})$. Furthermore,  we define the hyperedges $s_i''$ for $i\in I'$ as follows:
$$\quad s_i''\coloneqq\begin{cases}
    s_i, & \text{if }i\in I\,,\\
    \{ a, i \}, & \text{if }i\in X\,,\\
    X', & \text{if }i=i_b\,,\\
    \{ a \}, & \text{if }i=i_c\,.\\
\end{cases}$$ 
 Define $f': X'\to \{0,1,2\}$ with $f'^{-1}(2)=U\cup \{a\}$ and $f^{-1}(0)=X\setminus U$. Thus $\vert f'^{-1}(1)\vert =0$, i.e., the parameter value is constant.
\begin{claim}
    $(H,U)$ is a \yes-instance of \EHS \iffl $(H'',\tau,f')$ is a \yes-instance of \ExtRHF.
\end{claim}
\begin{pfclaim}
First, we assume that there is a minimal \rhf $g$ on $H'', \tau$ with $f' \leq g$. Define $D\coloneqq g^{-1}(2)\setminus \{a\}$. As $\tau^{-1}(i)=\emptyset$ for each $i\in I$, $s_i\cap D$ is not empty and $D$ is a \hs of~$H$. If there exists an $x\in D$ such that, for each $i\in \mathbf{I}(x)$, there is a $y\in s_i\cap (D\setminus \{x\})$, then this would contradict Constraint \ref{con_12switch} of \autoref{t_property_min_rhf}. Hence, $D$ is a minimal \hs. 

Assume that there is a minimal \hs $D\subseteq X$ on $H$ with $U\subseteq D$. Define $g:X\to \{0,1,2\}$ with $g^{-1}(2) = D \cup \{a\}$ and $g^{-1}(0) = X \setminus D$. Trivially, $f'\leq g$ holds. We will use \autoref{t_property_min_rhf} to show that $g$ is a minimal \rhf. Since $g^{-1}(1)$ is empty, the first two constraints, i.e., Constraints~\ref{con_difvalue} and~\ref{con_disjoint}, are fulfilled. As $D$ is a minimal \hs of $H$ and each hyperedge of $H$ is also a hyperedge of $H'$, Constraint \ref{con_12switch} also holds. For the last constraint, we have to consider $i\in I$, only, since $g(a)=2$. From the fact that $D$ is a minimal \hs on $H$ and $s''_{i_c}$ is only hit by~$a$, the last constraint of \autoref{t_property_min_rhf} follows. Thus, $g$ is a minimal \rhf.
\end{pfclaim}

\noindent
This finishes the discussion of all parameters.
\end{pf}
\end{toappendix}
\shortversion{\noindent We can use one of the reductions of \autoref{t_paraNP_EXTRHF} to prove transversal-hardness. %The proof needs some more details which can be find in the appendix.
}

\noindent
Finally, we present some (rather straightforward) algorithmic results, based on different parameterizations.

\begin{theorem}\label{t_FPT_EXTRHF}
$\kappa_\zeta$-$\ExtRHF\in\FPT $ for \shortversion{\\}$\zeta\in\{\weightfunction{2-f},  \vert f^{-1}(\{0,1\}) \vert \}$.
\end{theorem}

\begin{pf} \shortversion{(Sketch)}\longversion{We will present a simple \FPT-algorithm.} Let $(H=(X,\hat{S}),\tau,f)$ be an instance. The idea is to walk through all functions $g:X\to \{0,1,2\}$ with $f\leq g$ and test if~$g$ is a minimal \rhf. This runs in \FPT-time\longversion{ as, for vertices $x\in X$ with $f(x)=0$, there are~3 choices, for $f(x)=1$, there are~2 choices and for $f(x)=2$, there is only~1 choice}. Furthermore, we can check in polynomial time  if a function is a minimal \rhf (\longversion{such an algorithm can be constructed }by modifying \autoref{alg:tau_sur}\longversion{ in a straightforward way}). \longversion{Therefore, there are  
\begin{eqnarray*}\left( \prod_{x\in  f^{-1}(0)}3\right) \cdot \left( \prod_{x\in  f^{-1}(1)}2 \right)&\leq& \left( \prod_{x\in  f^{-1}(0)}2^{2-f(x)}\right) \cdot \left( \prod_{x\in  f^{-1}(1)}2^{2-f(x)} \right)\\& =& 2^{\weightfunction{2-f(x)}}\end{eqnarray*}
or
\begin{eqnarray*}\left( \prod_{x\in  f^{-1}(0)}3\right) \cdot \left( \prod_{x\in  f^{-1}(1)}2 \right) &\leq& \left( \prod_{x\in  f^{-1}(0)}3\right) \cdot \left( \prod_{x\in  f^{-1}(1)}3 \right)\\&=& 3^{\vert f^{-1}(\{0,1\})\vert } \end{eqnarray*}
many possibilities for $g$.}\shortversion{Hence, there are $2^{\weightfunction{2-f(x)}}$ or  $3^{\vert f^{-1}(\{0,1\})\vert }$ many possibilities for $g$.}
\end{pf}

\section{Enumerating Minimal Roman Hitting Functions}
\label{sec:enumerating_rhf}

In %\reviewer{There are several notations that should be recalled (or the reader should be guided toward the relevant sections) if the authors care about the enumeration section to be readable without having read all previous hardness results. These include the operations on functions that you use in Observation 1, notation mu-cal{RHF}, notation V1 and V2.} 
this section, we consider the output-sensitive complexity of enumerating minimal \rhfs. We start by showing that enumerating minimal \rhfs is transversal-hard. \amendment{We then complement the transversal-hardness result by proving the converse implication: if all minimal hitting sets of an arbitrary hypergraph can be enumerated with polynomial delay, then all minimal Roman hitting functions of an arbitrary instance $(H, \tau)$ can be enumerated with polynomial delay. Together with the transversal-hardness theorem, this establishes a polynomial-delay equivalence between the two enumeration problems. Additionally, the technique we use in this converse direction allows us to prove, with some modification and additional analysis, that there is an \FPT-delay algorithm to enumerate the minimal \rhfs of an arbitrary instance $(H, \tau)$ parameterized by  $\kappa_1$.} 

From these results, it \clarification{follows as a corollary} that enumerating minimal \rhfs is possible with polynomial delay if the correspondence function is surjective.

\amendment{Recall that for a given hypergraph $H$ and correspondence function $\tau$, $\mRhfSet{H}$ denotes the set of minimal \rhfs with respect to $H$ and $\tau$.}

\begin{theorem}[Transversal hardness theorem]
\label{t_transversal_hardness_rhf}
    If there would be an algorithm to enumerate all minimal \rhf of an \clarification{arbitrary} instance $(H,\tau)$ (even when $\tau$ is injective) with polynomial delay, then there is an algorithm that enumerates all minimal hitting sets of \clarification{an arbitrary} hypergraph $H'$ with polynomial delay. %\todo{Ich denke wir können das mit mit transversal hard verkürzen.}\todo{Wir haben Transversal hard tatsächlich schon eingeführt.}
\end{theorem}

\begin{pf}
    Let %\reviewer{The proof relies on the construction given in the proof of Theorem 13, which in turn relies on some notations that are not recalled: U is not defined, neither are a, ib, ic, etc. It is unfortunate that such a long and technical paper cannot be read partially. That being said, I have the feeling that theorem 15 does not actually deserve a construction nor a proof. Isn't it true that Roman hitting functions behave like regular hitting sets once we the correspondance function is empty? If true I believe it is worth making this observation after the definition of Roman hitting functions at the beginning of the paper.}
    $H=(X,(s_i)_{i\in I})$ be a hypergraph. We will use the same construction for a hypergraph $H''=(X',\{s_i''\}_{i\in I'})$ as in the proof of \autoref{t_paraNP_EXTRHF}.  Recall the following notation: $X' := \{a\} \cup X$, $I' := \{i_b,i_c\} \cup X \cup I$, $$\tau': X'\to I', x\mapsto \begin{cases}
            x, & x\in X\\
            i_b, & x=a
    \end{cases}, \quad \forall i\in I'\colon s_i''\coloneqq\begin{cases}
    s_i, & \text{if }i\in I\,,\\
    \{ a, i \}, & \text{if }i\in X\,,\\
    X', & \text{if }i=i_b\,,\\
    \{ a \}, & \text{if }i=i_c\,.\\
\end{cases}$$ 

    %As recall we define $X'=X \cup \{a\}$ and $I''= I\cup \{i_b,i_c\}$ as well as the hypergraph $H''= (X',(s_i'')_{i\in I''})$. Furthermore, we need 
%$$\tau': X'\to I'', x\mapsto \begin{cases}
%    x, & x\in X\\
%    i_b, & x=a
%\end{cases}, \quad s_i''\coloneqq\begin{cases}
%    s_i, & i\in I\\
%    \{ a, i \}, & i\in X\\
%    X', & i=i_b\\
%    \{ a \}, & i=i_c\\
%\end{cases}$$ 
%for $i\in I''$.
Since $\tau^{-1}(i_c) = \emptyset$ and $s''_{i_c}=\{a\}$, $f(a)=2$ for each \rhf $f:X'\to \{ 0, 1, 2\}$. Thus, for each minimal \rhf $f:X'\to \{ 0, 1, 2\}$ and $x\in X$, $f(x)\neq 1$, otherwise it contradicts Constraint \ref{con_disjoint} of \autoref{t_property_min_rhf}. Hence, $f^{-1}(1)$ is empty for each minimal \rhf $f$ of~$H''$.

We will show that there is a bijective function~$T$ between the minimal hitting sets of~$H$ and the minimal \rhf of~$H''$. This function maps each minimal \hs~$D$ to $g:X'\to \{0,1,2\} $ with $g^{-1}(0)= X\setminus D$, $g^{-1}(2)=\{a\}\cup D$. In the proof of \autoref{t_paraNP_EXTRHF}, we  have already shown that $g$ is a minimal \rhf of~$H''$. The inverse function maps a minimal \rhf~$g$ to $g^{-1}(2)\setminus \{a\}$, which is a minimal \hs by the proof of \autoref{t_paraNP_EXTRHF}. Since we can compute $T(D)$ and $T^{-1}(g)$ for a minimal \hs and for a minimal \rhf~$g$, respectively, in polynomial time, the theorem holds.   
\end{pf}

\amendment{We now define some terminology which the subsequent theorems in this section use. An index $i \in I$ is \emph{owned} if $\tau^{-1}(i)\neq\emptyset$, and \emph{unowned} otherwise. We write
$I_{\mathrm{own}}\coloneqq\tau(X)$ and
$I_{\mathrm{un}}\coloneqq I\setminus\tau(X)$.}

For a hypergraph $H=((X,(s_i)_{i \in I})$ and a correspondence function  $\tau$, define $P_{H,C}(x)\coloneqq \mathbf{I}(x)\setminus \mathbf{I}(C\setminus \{x\})$ and %\reviewer{Provide intuition for P(H,tau).}
$$\reformulation{\mathcal{R}(H,\tau)}\coloneqq \{C\subseteq X\mid \{i\in I\mid \tau^{-1}(i)=\emptyset\}\subseteq \mathbf{I}(C) \wedge  \forall x\in C:\: P_{H,C}(x)\nsubseteq \{\tau(x)\}\}.$$ 

\clarification{The set $P_{H,C}(x)$ consists of the indices of the
hyperedges that are private to $x$ with respect to $C$; that is, these edges are hit by $C$ only at $x$.} \amendment{We call every set
$C\in \reformulation{\mathcal{R}(H,\tau)}$ a \emph{Roman support} of $(H,\tau)$.
Equivalently, a Roman support hits every unowned hyperedge, and each
of its vertices has a private hyperedge whose index is different
from the index owned by that vertex. Here, ``support'' refers
specifically to the set $f^{-1}(2)$, rather than to the ordinary
support $f^{-1}(\{1,2\})$ of a function. \autoref{cor:rhf_existence_set}, which follows from the lemma below, tells us that a Roman support is a necessary support component of a minimal \rhf.}

\begin{lemma}\label{lem:sur_rhf_to_R}
    Let $H=(X,(s_i)_{i \in I})$ be a hypergraph and $\tau$ be a correspondence function. Then $\Psi_H: \mRhfSet{H} \to \reformulation{\mathcal{R}(H,\tau)}$, $f \mapsto f^{-1}(2)$ is a surjective function, and for all $C\in \reformulation{\mathcal{R}(H,\tau )}$, %\reviewer{$\Psi^{-1}$ should probably be $\Psi_H^{-1}$}
    $\Psi_H^{-1}(C)$ can be enumerated with polynomial delay and polynomial space.
\end{lemma}

\begin{proof}
    By \autoref{t_property_min_rhf}, $\Psi_H$ is well-defined. Let $C\in \reformulation{\mathcal{R}(H,\tau)}$.  Let $(x_i)_{i\in I\setminus \mathbf{I}(C)}$ be a sequence such that, for all $i\in I\setminus \mathbf{I}(C)$, $x_i\in \tau^{-1}(i)$. Define $f_C \coloneqq 2\chi_{C}+ \chi_{\{x_i\mid i\in I\setminus \mathbf{I}(C)\}}$. $f_{C}$ is minimal by \autoref{t_property_min_rhf} as, for each $i\in I\setminus \mathbf{I}(C)$, there is exactly one element $x_i\in f_C^{-1}(1)$. Hence, $\Psi_H$ is surjective. 

    Now, we want to prove that for all $C\in \reformulation{\mathcal{R}(H,\tau)}$ and $f\in \Psi_H^{-1}(C)$, there exists a sequence $(x_i)_{i \in I\setminus \mathbf{I}(C)}$ with $x_i\in \tau^{-1}(i)$ for all  $i\in I\setminus \mathbf{I}(C)$ such that $f = 2\chi_{C}+ \chi_{\{x_i\mid i\in I\setminus \mathbf{I}(C)\}}.$ Since $f$ is a \rhf, there exists  a $y_i \in \tau^{-1}(i)$ for each $i\in I\setminus\mathbf{I}(C)$. Thus, $f'= 2\chi_{C} + \chi_{\{y_i\mid i\in I\setminus \mathbf{I}(C)\}}\leq f$. Clearly, $f'$ is a \rhf. Therefore, $f=f'$ and there exists a sequence $(y_i)_{i\in I\setminus\mathbf{I}(C)}$ with $f = 2\chi_{C}+ \chi_{\{y_i\mid i\in I\setminus \mathbf{I}(C)\}}.$ Keep in mind that we can choose any combination of elements from $\tau^{-1}(i)$ for all $i\in I\setminus \mathbf{I}(C)$. The next combination can therefore be constructed in polynomial time and at most quadratic space. Thus, we can enumerate $\Psi_H^{-1}(C)$ with polynomial delay and polynomial space.  
\end{proof}

\noindent
Since $\Psi_H$ is a well-defined surjective mapping, we obtain the following corollary.

\begin{corollary}\label{cor:rhf_existence_set}
    Let $H=(x,(s_i)_{i\in I})$ be a hypergraph, $\tau$ be a correspondence function and $C \subseteq X$. Then, there exists a minimal \rhf $f\in \{0,1,2\}^X$ with $f^{-1}(2)=C$ \iffl $C\in \reformulation{\mathcal{R}(H,\tau)}$. 
\end{corollary}

\begingroup
\begin{theorem}\label{t_converse_transversal_hardness}
If the minimal hitting sets of arbitrary hypergraphs can be
enumerated with polynomial delay, then the minimal \rhfs of every
instance $(H,\tau)$ can be enumerated with polynomial delay.
\end{theorem}

\begin{pf}

Fix an arbitrary hypergraph $H=(X,(s_i)_{i\in I})$ and an arbitrary
correspondence $\tau:X\to I$. Define $H_{\mathrm{un}}
\coloneqq
(X,(s_i)_{i\in I_{\mathrm{un}}})$.

\begin{claim}[Deletion property]
Let $B$ be a Roman support and let $C\subseteq B$. If $C$ hits every
unowned hyperedge, then $C$ is a Roman support.
\end{claim}

\begin{pfclaim}
By assumption, $C$ hits every unowned hyperedge. In order to certify the additional private hyperedge condition necessary to render a set a Roman support, let $x \in C$ be arbitrary. Then $x \in B$, and by virtue of $B$ being a Roman support, there is some index  $i_x \in I \setminus \{\tau(x)\}$ for which $  s_{i_x}\cap B=\{x\}$. From $C\subseteq B$ and $x\in C$, it follows that $ s_{i_x}\cap C=\{x\}$. $i_x\neq\tau(x)$ holds as a simple consequence of the nature of the set over which
$i_x$ is quantified. Therefore, $C$ meets all of the criteria required to be a Roman support.
\end{pfclaim}

\begin{claim}
The minimal hitting sets of $H_{\mathrm{un}}$ are exactly the
inclusion-minimal Roman supports of $(H,\tau)$ in $\mathcal{R}(H,\tau)$ (a Roman support $ B \in \mathcal{R}(H,\tau)$  is inclusion-minimal in $\mathcal{R}(H,\tau)$ if there is no $C \in \mathcal{R}(H,\tau)$ for which $C \subsetneq B)$. 
\end{claim}

\begin{pfclaim}
We prove that the set of all minimal hitting sets of $H_{\mathrm{un}}$ is equal to the set of all inclusion-minimal Roman supports of $(H,\tau)$ by showing both of the relevant inclusions. 

For the forward inclusion, let $T$ be a minimal hitting set of $H_{\mathrm{un}}$. Since $T$ is a hitting set of $H_{\mathrm{un}}$, it hits every
hyperedge $s_i$ with $i\in I_{\mathrm{un}}$. Now let $x \in T$ be arbitrary. By the minimality of $T$, $T \setminus \{x\}$ fails to hit some unowned hyperedge $s_i$ with $i \in I_{\mathrm{un}}$. Now we have $s_i \cap T = \{x\}$: $x$ must be part of the intersection, as the fact that it was the specific absence of $x$ which made $T \setminus \{x\}$ fail to hit $s_i$ implies $x \in s_i$; moreover, $s_i \cap T$ cannot contain elements other than $x$, as then $T \setminus \{x\}$ would have indeed hit $s_i$. Crucially, $i\in I_{\mathrm{un}}=I\setminus\tau(X)$, so we have $i\neq\tau(x)$. Thus $i\in P_{H,T}(x)\setminus\{\tau(x)\}$, and $T$ is a Roman support. $T$ must be an inclusion-minimal Roman support, because any proper subset of $T$ would fail to hit  some hyperedge of
$H_{\mathrm{un}}$.

Now let $B$ be an inclusion-minimal Roman support.  By definition,
$B$ is a hitting set of $H_{\mathrm{un}}$. Assume for the sake of
contradiction that $B$ is not a minimal hitting set of
$H_{\mathrm{un}}$. Then there exists a set $C\subsetneq B$ that is a
hitting set of $H_{\mathrm{un}}$. By the previous deletion claim, $C$ is a Roman support, which contradicts the inclusion-minimality of
$B$. Hence, $B$ is a minimal hitting set of $H_{\mathrm{un}}$. \end{pfclaim}

Fix an arbitrary total order $\prec$ on $X$. For every $B$ in $\mathcal{R}(H,\tau)$, define \[D(B)
    \coloneqq
    \{x\in B\mid B\setminus\{x\}\in \mathcal{R}(H,\tau)\}\]
Thus, $D(B)$ is the set of vertices whose deletion from $B$ leaves
another Roman support.

If $D(B)\neq\emptyset$, define the parent of $B$ by
\[
    p(B)
    \coloneqq
    B\setminus\{\max_{\prec}D(B)\}.
\]

Since $X$ is finite and $\prec$ is a total order on $X$, $\max_{\prec}D(B)$ is uniquely determined whenever $D(B)$ is nonempty. If $D(B)  = \emptyset$, then we call $B$ a root. 

Let $\mathcal{F}_p$ be the directed graph with vertex set
$\mathcal{R}(H,\tau)$ and with an arc from $B$ to $p(B)$ for every
$B\in \mathcal{R}(H,\tau)$ satisfying $D(B)\neq\emptyset$.
In the claim  below, a rooted forest is an acyclic graph whose constituent trees have one of their nodes designated as a root.

\begin{claim} The directed graph $\mathcal{F}_p$ is a rooted forest whose roots --- in the above sense of the elements of $\mathcal{R}(H,\tau)$ which render $D(\cdot)$ empty ---  are
exactly the minimal hitting sets of $H_{\mathrm{un}}$.
\end{claim}
\begin{pfclaim}

First we show that $D(B) = \emptyset$ if and only if $B$ is inclusion-minimal in $\mathcal{R}(H,\tau)$. By the previous claim,  this would imply that the roots of $\mathcal{F}_p$ are exactly the minimal hitting sets of $H_{\mathrm{un}}$.

If $D(B)\neq\emptyset$, then $B\setminus\{x\}\in \mathcal{R}(H,\tau)$ for some $x \in B$, and $B$ is not inclusion-minimal in $\mathcal{R}(H,\tau)$.

Conversely, assume that $B$ is not inclusion-minimal in $\mathcal{R}(H,\tau)$. Then there exists a $C \in \mathcal{R}(H,\tau)$ for which $C \subsetneq B$. Let $x \in B \setminus C$. Since $C$ is a Roman support, it hits every hyperedge of $H_{\mathrm{un}}$; this implies that $B \setminus \{x\}$ hits every hyperedge of $H_{\mathrm{un}}$, because $C \subseteq B \setminus \{x\}$. The deletion claim earlier in this proof then gives us that $B \setminus \{x\}$ is a Roman support. As such, $x \in D(B)$, and $D(B) \neq \emptyset$.

Now we show that $\mathcal{F}_p$ is a rooted forest. Every non-root Roman support $B$ has exactly one parent: $D(B) \neq \emptyset$, and the total order $\prec$ makes $p(B)$ unique. Moreover, by the definition of $D(B)$, $p(B)$ is itself a Roman support, and $|p(B)| = |B| - 1$. Since $\mathcal{R}(H,\tau) \subseteq 2^X$ is finite, repeated application of $p$ from any Roman support must terminate at a Roman support $B^*$ for which $D(B^*)=\emptyset$. Thus every Roman support belongs to a component containing a root. The root at which repeated application of $p$ must eventually terminate is unique, because the parent of any Roman support is unique. 

Moreover, it is impossible for there to be a cycle in $\mathcal{F}_p$: every directed arc in $\mathcal{F}_p$ leads from a Roman support to its parent, and thus the head of such an arc is a set with strictly lesser cardinality than the tail of said arc. The presence of a cycle in $\mathcal{F}_p$ would imply that a Roman support in $\mathcal{F}_p$ has two cardinalities --- ``its" cardinality, and then an even smaller cardinality based on applying the cardinality-decreasing property of successive arcs leading back to said Roman support.

Therefore, $\mathcal{F}_p$ is a rooted forest.\end{pfclaim}

We now use the rooted forest $\mathcal{F}_p$ to define a reverse-search algorithm, which algorithm uses as a tool the  polynomial-delay algorithm for enumerating the minimal hitting sets of $H_\mathrm{un}$ whose existence this proof assumes. When the minimal-hitting-set enumerator outputs a root, this is not itself an output of the reverse-search algorithm at that time; the root is only output by the reverse-search algorithm later when the recursive procedure it employs is invoked on it. 

The reverse-search algorithm performs the following steps. Run (simulate) the  polynomial-delay algorithm for enumerating the minimal hitting sets of $H_\mathrm{un}$ until it either outputs a minimal hitting set of $H_\mathrm{un}$ or reports that no further minimal hitting set exists. If the latter eventuality occurs, the reverse-search algorithm terminates. In the former case, the reverse-search algorithm suspends the minimal-hitting-set enumerator and runs the recursive procedure specified by the following on the enumerated root. The recursive procedure takes as input a Roman support $B$, and first outputs~$B$. Then, for each $y \in X \setminus B$, the procedure sets $C \coloneqq B \cup \{y\}$, and if $C$ is a Roman support and $p(C) = B$, the procedure recursively visits $C$. After the recursive procedure terminates, the reverse-search algorithm resumes the minimal-hitting-set enumerator, and repeats the same treatment of the assumed enumerator: if it outputs a root, the reverse-search algorithm pauses the minimal-hitting-set enumerator and runs the recursive procedure outlined above on the enumerated root, upon the conclusion of which it resumes the minimal-hitting-set enumerator; if the minimal-hitting-set enumerator terminates, the reverse-search algorithm terminates. This process continues until the minimal-hitting-set enumerator, and in turn the reverse-search algorithm, terminates.

\begin{comment}
\begin{enumerate}
\item Run the polynomial-delay algorithm for enumerating the minimal hitting sets of $H_\mathrm{un}$, whose existence this proof assumes. By the preceding claim, these sets are exactly the roots of $\mathcal{F}_p$. The outputs of the minimal-hitting-set enumerator are used as inputs to the recursive procedure defined below, and are not separately output by the overall reverse-search algorithm.
\item The roots are produced one-by-one by the previous step. For each root $R$, run the recursive procedure specified by the following on it. The recursive procedure takes as input a Roman support $B$, and first outputs $B$. Then, for each $y \in X \setminus B$, the procedure sets $C \coloneqq B \cup \{y\}$, and if $C$ is a Roman support and $p(C) = B$, the procedure recursively visits $C$.
\item When the hitting-set-enumerator reports that no further minimal hitting set exists, terminate.
\end{enumerate}
\end{comment}

For a vertex  $B$ of  $\mathcal{F}_p$, define the depth of $B$ to be the number of arcs in the unique directed path from $B$ to the root of its tree. 

\begin{claim}[Correctness of reverse search]
The reverse-search algorithm outputs every Roman support exactly once. 
\end{claim}

\begin{pfclaim}
We first show that every Roman support is output by the reverse-search algorithm. Fix a rooted tree of $\mathcal{F}_p$; we show that for all depths~$d$ in the tree, every Roman support in said tree at depth $d$ is output.  We proceed by induction on~$d$. 

At depth $0$, the root of the tree is a minimal hitting set of $H_\mathrm{un}$ by the preceding claim; it will thus be produced due to the assumed existence of a minimal-hitting-set enumerator, and the reverse-search algorithm will call its recursive procedure on it and output it. 

Now assume that the reverse-search algorithm outputs every Roman support at depth $d$ in the fixed tree. Let $C$ be a Roman support in the same tree at depth $d+1$. Then $p(C) = B$ for some roman support $B$ at depth $d$ in the tree. Since $p(C) = B$, there is a unique vertex $y \in X \setminus B$ for which $C = B \cup \{y\}$. By the induction hypothesis, the reverse-search algorithm outputs $B$. Therefore, it visits $B$, considers this vertex $y$, and, conditional on $C$ both being a Roman support and satisfying $p(C) = B$, outputs $C$. It is indeed the case that $C$ is a Roman support and $p(C) = B$. Thus, after validating that these two conditions hold, the reverse-search algorithm will recursively visit $C$, whereupon the set~$C$ will be output. 

We next show by induction on  $d$ that for every Roman support~$B$ at depth $d$ in a rooted tree of $\mathcal F_p$, the set~$B$ is output no more than once by the reverse-search algorithm.

If $d= 0$, this holds because every root is produced exactly once by the minimal-hitting-set enumerator, and the reverse-search algorithm invokes the recursive procedure exactly once on each root produced by that enumerator. Moreover, a sequence of recursive calls made by the recursive procedure cannot return to the root on which it was invoked, because the recursive procedure, when visiting a Roman support $B$, only recurses on Roman supports of which $B$ is the parent, and a root has no parent. 

Now assume that every Roman support of depth $d$ is output at most once. Then if $C$ is a Roman support of depth $d+1$, its unique parent $p(C)$ is output at most once, and is therefore visited by the recursive procedure at most once. When the recursive procedure is visiting $p(C)$, it will only consider $C$ once, because there is a unique vertex $y \in C \setminus p(C)$ such that $C = p(C) \cup \{y\}$, so the recursive procedure only considers $y$ once. Because a non-root Roman support can be output only when the reverse-search procedure is visiting its unique parent, the set~$C$ is output no more than once by the reverse-search algorithm. 
\end{pfclaim}

\begin{claim}
The reverse-search algorithm enumerates all Roman supports with polynomial delay.
\end{claim}

\begin{pfclaim}
By assumption, the roots of $\mathcal{F}_p$ (i.e., the minimal hitting sets of $H_\mathrm{un}$) can be enumerated with polynomial delay. Thus the time before the reverse-search algorithm outputs its first Roman support is polynomial.

We now argue that the time between two successive enumerations of the reverse-search algorithm, if both enumerations are elements in the same rooted tree of $\mathcal{F}_p$, is polynomially bounded. Between enumerating two elements in the same rooted tree of $\mathcal{F}_p$, the reverse-search algorithm will only check, for each subset $C$ of $X$ formed by adding to the set $B$ it is visiting some $y \in X \setminus B$, whether $C$ is a Roman support and $p(C) = B$, and if each of these checks come back in the negative, it will backtrack to some other Roman support. During this process, when the reverse-search algorithm is still visiting $B$, polynomially many such $C$ are considered hereby; in particular, the reverse-search algorithm constructs exactly  $|X \setminus B|$ of them. For a single set $C$ for which $|C| = |B| + 1$, performing the two checks which the reverse-search algorithm performs on $C$ consumes a polynomial amount of time. Namely, testing whether $C$ is a Roman support can be done in polynomial time: one can scan all indices in $I_{\mathrm{un}}$ to verify that $C \cap s_i \neq \emptyset$ for each $i \in I_{\mathrm{un}}$, and, for every $x \in C$, scan all indices $i\in I\setminus\{\tau(x)\}$ and check whether there exists at least one such index satisfying $s_i\cap C=\{x\}$. The check $p(C) = B$ can also  be done in polynomial time. $D(C)$ can be computed by testing, for every $c \in C$, whether $C\setminus\{c\}\in\mathcal{R}(H,\tau)$; and if $D(C) \neq \emptyset$, computing $\max_{\prec}D(C)$ and comparing $C\setminus\{\max_{\prec}D(C)\}$ with $B$ completes the desired check. If the reverse-search algorithm  backtracks to another Roman support after performing these checks, it will do so in polynomial time: the depth of a rooted tree in $\mathcal{F}_p$ is at most $|X|$, since the cardinality of the vertices in a directed path in $\mathcal{F}_p$ decreases by $1$ across every directed arc, due to the nature of the parent function. 

After the reverse-search algorithm has finished enumerating each of the Roman supports in one rooted tree, the polynomial-delay minimal-hitting-set enumerator whose existence we are assuming produces the next root, or reports that no root remains, in polynomial time. Therefore, the reverse-search algorithm enumerates all Roman supports with polynomial delay. 
\end{pfclaim}

The enumeration algorithm $A_\mathrm{enum}$ we give for the minimal \rhfs uses the reverse-search algorithm as a tool. As with the reverse-search algorithm's use of the minimal-hitting-set enumerator, $A_\mathrm{enum}$ runs  the reverse-search algorithm internally, pausing it after it outputs a Roman support. Specifically, whenever the reverse-search
algorithm produces a Roman support $B$, $A_\mathrm{enum}$ suspends the reverse-search algorithm and enumerates all minimal \rhfs~$f$ satisfying $f^{-1}(2)=B$ using the polynomial delay procedure from \autoref{lem:sur_rhf_to_R}. $A_\mathrm{enum}$ then resumes the simulation of the reverse-search algorithm until it either outputs a Roman support or terminates. When the reverse-search algorithm terminates, $A_\mathrm{enum}$ terminates. 

The outputs of the reverse-search algorithm which ensue due to its use as an internal tool of $A_\mathrm{enum}$ are not actually enumerated by $A_\mathrm{enum}$.

First of all, $A_\mathrm{enum}$ outputs every minimal \rhf.  To see this, let $f$ be an arbitrary minimal \rhf. By \autoref{lem:sur_rhf_to_R}, $f^{-1}(2)\in\mathcal{R}(H,\tau)$. Thus $f^{-1}(2)$ is a Roman support, and the reverse-search algorithm will eventually output $B$ such that $B=f^{-1}(2)$. At this point, $A_\mathrm{enum}$ employs the procedure from \autoref{lem:sur_rhf_to_R} to enumerate all minimal \rhfs~$g$ satisfying $g^{-1}(2)=B$; in particular $f$ is output by $A_\mathrm{enum}$.

Now suppose for the sake of contradiction that $A_\mathrm{enum}$ enumerates some minimal \rhf~$f$ twice. The two enumerations of $f$ on the part of $A_\mathrm{enum}$ could not have occurred when invoking the polynomial delay procedure from \autoref{lem:sur_rhf_to_R} to enumerate all minimal \rhfs satisfying $f^{-1}(2)=B$, since,  if they had done so, the referenced polynomial delay procedure from \autoref{lem:sur_rhf_to_R} would not be a valid enumeration algorithm. Therefore, there must exist two Roman supports $B_1$ and $B_2$ which the reverse-search algorithm enumerates at different times, for which $f^{-1}(2)=B_1$ and $f^{-1}(2)=B_2$. But then $B_1 = B_2$ holds by transitivity, and $B_1$ and $B_2$ could not have been output at different times by the reverse-search algorithm.

The algorithm $A_\mathrm{enum}$ has polynomial delay. If the reverse-search algorithm reports that no Roman support
exists, then $A_{\mathrm{enum}}$ terminates within polynomial time
without producing an output. Otherwise, the time between the initialization of $A_\mathrm{enum}$ and the first output of $A_\mathrm{enum}$ is polynomial, because the reverse-search algorithm enumerates its first output a polynomial amount of time after its initiation, and the enumerator from \autoref{lem:sur_rhf_to_R} which $A_\mathrm{enum}$ uses as a subroutine likewise produces its first output in polynomial time. 

It follows from \autoref{lem:sur_rhf_to_R} that the delay between  two enumerations of minimal \rhfs satisfying $f^{-1}(2) = B$, for some Roman support $B$, is polynomial. By the surjectivity of $\Psi_H$ established by \autoref{lem:sur_rhf_to_R}, for every Roman support $B$, there exists a minimal \rhf~$f$ for which $f^{-1}(2) = B$. Thus, after the reverse-search algorithm outputs a Roman support, $A_\mathrm{enum}$ enumerates a minimal \rhf within a polynomial amount of time. Between the last function in the preimage of one Roman support and the first function in the preimage of the next Roman support, there is but a polynomial-delay termination of the enumerator from \autoref{lem:sur_rhf_to_R}, a polynomial-delay transition to the next Roman support taken up by reverse-search, and a polynomial-delay initialization of the enumerator from \autoref{lem:sur_rhf_to_R}, in that order. Finally, the time between the last enumerated minimal \rhf and the termination of $A_\mathrm{enum}$ is polynomial, because the time between the last enumerated minimal \rhf and the termination of the enumerator from \autoref{lem:sur_rhf_to_R} is polynomial, and the time between last enumerated Roman support of the reverse-search algorithm and the termination of the reverse-search algorithm is polynomial. This proves the theorem. \end{pf}

\noindent
Theorems~\ref{t_transversal_hardness_rhf} and  \ref{t_converse_transversal_hardness} give us the following corollary. 
\begin{corollary}[Polynomial-delay Equivalence]

There exists a polynomial-delay algorithm that, given an arbitrary instance $(H,\tau)$, enumerates all of the minimal \rhfs of  $(H,\tau)$, if and only if there exists a polynomial-delay algorithm that, given an arbitrary hypergraph, enumerates all of its minimal hitting sets.
\end{corollary}
\endgroup

\reformulation{Since there is no known polynomial-delay algorithm for enumerating minimal hitting sets, we now consider a corresponding parameterized enumeration problem in the following. 
The notation is based on \cite{CreMMSV2017}.}

$E=(Q,\kappa,\text{Sol})$ is a parameterized enumeration problem (over a finite alphabet $\Sigma$) if 
\begin{itemize}
    \item $Q\subseteq \Sigma^*$,
    \item $\kappa \colon \Sigma^* \to \mathbb{N}$, the so-called parameterization, is a computable function, and
    \item $\text{Sol}: \Sigma^* \to 2^{\Sigma^*}$ is a function such that for all $x\in \Sigma^*$, $\text{Sol}(x)$ is a finite set and $\text{Sol}(x) \neq \emptyset$ \iffl $x\in Q$. 
\end{itemize}

For a parameterized enumeration problem $E=(Q,\kappa,\text{Sol})$ over the alphabet $\Sigma$, $x\in \Sigma^*$ is an instance, with the parameter $\kappa(x)$, and $\text{Sol}(x)$ is the set of solutions of $x$ (for an appropriate notion of `solution', e.g., a minimal hitting set). The goal is to enumerate all elements in $\text{Sol}(x)$ without repetition.

An enumeration algorithm for  a parameterized enumeration problem has \FPT-delay if there exists a polynomial $p$ and a computable function $t:\mathbb{N}\to \mathbb{N}$ such that the time from the start of the program to outputting the first solution, and the time from outputting  the last output to the program halt, as well as the time between any two consecutive solution outputs is bounded by $t(\kappa(x))p(\vert x\vert)$. 

\begingroup

The following corollary is obtained by a combination of the proof of \autoref{t_converse_transversal_hardness} and the minimal-hitting-set enumeration algorithm of \cite{BlaFLMS2022}. 

\begin{corollary}[FPT-delay parameterized by $\kappa_1$]

There exists a polynomial-space, \FPT-delay enumeration algorithm for minimal \rhfs when parameterized by \(\kappa_1 = |I_\mathrm{un}|\).
\end{corollary}
\begin{pf}
Let $H = (X,(s_i)_{i\in I})$ be a hypergraph and $\tau: X \to I$ a correspondence function. As in the proof of \autoref{t_converse_transversal_hardness}, define $H_{\mathrm{un}}
\coloneqq
(X,(s_i)_{i\in I_{\mathrm{un}}})$. Because we would like to apply the minimal-hitting-set enumerator from~\cite{BlaFLMS2022}, but that minimal-hitting-set enumerator applies only to hypergraphs whose edges collection is a set, in contradistinction to how  $H_{\mathrm{un}}$ may contain repeated hyperedges, we define an auxiliary hypergraph formed by deleting repeated hyperedges in $H_{\mathrm{un}}$. Let $\mathcal S_\mathrm{un} = \{ s_i \mid i \in I_\mathrm{un}\}$ and define $\overline{H}_\mathrm{un} =  (X, \mathcal S_\mathrm{un})$. Because repeated hyperedges do not affect what the minimal hitting sets are, enumerating the minimal hitting sets of $\overline{H}_\mathrm{un}$ is equivalent to enumerating the minimal hitting sets of~$H_\mathrm{un}$.

If $\emptyset \in \mathcal S_\mathrm{un}$, then $\overline{H}_\mathrm{un}$ lacks a hitting set, so there is neither a Roman support nor minimal \RHF; this case can be trivially handled by defining a minimal-\RHF enumerator with the properties desired by this Corollary which terminates without producing an output. Therefore, we assume for the rest of this proof that $\emptyset \notin \mathcal S_\mathrm{un}$.

This assumption allows us to define $k^*$ as being the maximum cardinality of a minimal hitting set of $\overline{H}_\mathrm{un}$. We first show that \(k^* \le \kappa_1\). This is important because the delay bound which \cite{BlaFLMS2022} gives for its algorithm to enumerate the minimal hitting sets of an arbitrary hypergraph depends on both the number of hyperedges of said hypergraph and the maximum cardinality $k^*$ of a minimal hitting set of the same hypergraph. Our goal is to show that the minimal-hitting-set enumerator from \cite{BlaFLMS2022} has \FPT-delay parameterized by $\kappa_1$, so that we can use it in lieu of the assumed minimal-hitting-set enumerator in $A_\mathrm{enum}$ from the proof of \autoref{t_converse_transversal_hardness}.

Let $T$ be a minimal hitting set of $\overline{H}_\mathrm{un}$. Because $T$ is minimal, for every $x \in T$, there exists a hyperedge $e_x \in \mathcal S_\mathrm{un}$ for which $e_x \cap T = \{x\}$ (if there were no such hyperedge --- meaning that for every hyperedge of $\overline{H}_\mathrm{un}$, its intersection with $T$, if it contains $x$, must contain another element besides $x$ --- then $T \setminus \{x\}$ would be a hitting set of $\overline{H}_\mathrm{un}$, contradicting the minimality of $T$). Let  $x \mapsto e_x $  be any function from $T$ to $\mathcal S_\mathrm{un}$ which maps a vertex $x \in T$ to some hyperedge $e_x \in \mathcal S_\mathrm{un}$ for which $e_x \cap T = \{x\}$. $x \mapsto e_x$ is injective: if $x \neq y$, then if $e_x = e_y$, there is a single hyperedge $e_x$ which satisfies both $ e_x \cap T=\{x\}$ and $ e_x \cap T=\{y\}$, which is impossible. Therefore the codomain of $x \mapsto e_x$ has cardinality at least that of the domain of $x \mapsto e_x$; i.e., $|T| \le |\mathcal{S}_{\mathrm{un}} |\le \kappa_1$. Because $T$ was arbitrary, this gives $k^* \le \kappa_1$, as desired.  

By~\cite{BlaFLMS2022}, the minimal hitting sets of a hypergraph with $n$ vertices and $m$ hyperedges can be enumerated in a way which both consumes polynomial space and has delay $\Oh(m^{k^* + 1}n^2)$ . We express the delay of the minimal-hitting-set enumerator given by~\cite{BlaFLMS2022} as $\Oh(1+ m^{k^* + 1}n^2)$ in order to handle the degenerate edge case. Applying the given minimal-hitting-set enumerator to $\overline{H}_\mathrm{un}$ with $n = |X|$, $m = |\mathcal{S}_{\mathrm{un}}| \le \kappa_1$, and $k^* \le \kappa_1$ gives an algorithm for enumerating the minimal hitting sets of $\overline{H}_\mathrm{un}$ with $\Oh(1+ \kappa_1^{\kappa_1 + 1} |X|^2)$ delay and polynomial space. This delay can be written as $g(\kappa_1) \cdot|(H, \tau)|^{\Oh(1)}$ for the computable function $g(k) = 1 + k^{k+1}$, where $|(H, \tau)|$ denotes the length of the encoding for the instance $(H, \tau)$. Thus there exists a polynomial-space minimal-hitting-set enumerator with \FPT-delay, parameterized by $\kappa_1$.

Now consider the algorithm $A'_\mathrm{enum}$, which is the same as $A_\mathrm{enum}$ from the proof of \autoref{t_converse_transversal_hardness} in all respects except that its invocation of the reverse-search algorithm involves the reverse-search algorithm specifically using a polynomial-space minimal-hitting-set enumerator with \FPT-delay parameterized by $\kappa_1$ to generate the minimal hitting sets of $\overline{H}_\mathrm{un}$, rather than using an arbitrary minimal-hitting-set enumerator to enumerate the minimal hitting sets of $H_\mathrm{un}$. Because enumerating the minimal hitting sets of $\overline{H}_\mathrm{un}$ is equivalent to enumerating those of $H_\mathrm{un}$ per the discussion in the first paragraph of this proof, the correctness proof of $A_\mathrm{enum}$ applies to $A'_\mathrm{enum}$.

The reverse-search algorithm modified thus is a polynomial-space enumeration algorithm with respect to~$\kappa_1$ for all Roman supports. The modified reverse-search algorithm needs only space to run a polynomial-space minimal-hitting-set enumerator and to maintain its recursion stack of depth $|X|$, each of the frames of which store only a polynomial-size of information, namely the Roman support being visited, the next candidate vertex to be examined, and the data required by the membership and parent tests. 

Because the procedure from \autoref{lem:sur_rhf_to_R} is also polynomial-space, it follows that $A'_\mathrm{enum}$ is itself using only polynomial space. 
The algorithm $A'_\mathrm{enum}$ has \FPT-delay, because the delay between any two of its outputs is at most the sum of the \FPT-delays on the part of the minimal-hitting-set enumerator and the sum of the polynomial-delay operations which $A_\mathrm{enum}$ performs between any two of its outputs. \end{pf}

\endgroup

As $\kappa_1$ measures the `distance from surjectivity' of correspondence functions, we can hence deduce the following result that can be seen as a counterpart of \autoref{cor:ExtRHF-surjective} in the world of extension problems.

\begin{corollary}\label{cor:poly_delay_tau_sur}
    There exists a polynomial-delay and polynomial-space enumeration algorithm for minimal \rhfs with surjective correspondence functions. 
\end{corollary}

\section{Conclusions}

We have generalized the notion of Roman domination towards hypergraphs by introducing the definition of \rhfs. We have proven that all minimal \rhfs can be enumerated with polynomial delay if the correspondence function is surjective. This can be seen as a technical answer to our question what causes Roman domination to behave different from classical domination with respect to polynomial-delay enumerability. When the correspondence is not surjective, \RHF rather behaves like \DS; in particular, its extension problem is $\W{3}$-complete when parameterized by the given pre-solution's weight, and we observe that it is transversal-hard to enumerate all minimal \rhfs.

    We mentioned in the introduction that \RD is in \FPT, when parameterized in a dual way, meaning, in this case, by $n-k$, where $n$ is the number of vertices of the graph and $k$ is an upper-bound on the weight of the \rdf. It might be interesting to have similar results for the two generalizations of Roman domination introduced in this paper. However, now it is not very clear what the `dual' of the $\weightfunction{f}$-parameterization should be.
%\end{itemize}

%We are currently looking for non-trivial graph-classes where \bERD is 
%solvable in polynomial time, hence looking onto another tractability frontier.\todo{Are we still looking here? Or should we remove this paragraph?}

Notice that up to quite recently, only a handful of (natural) problems have been known to be complete for $\W{3}$\amendment{, which is the third level of the \textsf{W}-hierarchy of parameterized complexity}. Even today, apart from the extension problems and their relatives that we mentioned throughout this paper, we only know of the problems shown in \cite{BodGroPil2022,CheZha2006,CheFluGro2007}. Seeing more and more problems these days that are complete for $\W{3}$ adds new interest to~\W{3}. It might be time to attack the 25-years-old open question if $\W{3}=\textsf{W}^*[3]$, see \cite{DowFel98}. According to~\cite{CheFluGro2007}, the current status is: $\W{3}\subseteq\textsf{W}^*[3]\subseteq\W{4}$.

\bibliographystyle{splncs04}
\bibliography{ab,hen}

\end{document}